\documentclass[leqno]{siamltex}

\usepackage{amsmath,amssymb}
\usepackage{graphicx}
\usepackage{parskip}
\usepackage[pdfpagemode=UseNone,pdfstartview=FitH]{hyperref}
\usepackage{mathtools}  
\usepackage{stmaryrd}
\usepackage{doi}
\usepackage{array}
\usepackage{pstricks}
\usepackage{mathtools}
\usepackage{vmargin}
\usepackage{colortbl,xcolor}

\makeatletter
\DeclareRobustCommand\widecheck[1]{{\mathpalette\@widecheck{#1}}}
\def\@widecheck#1#2{%
    \setbox\z@\hbox{\m@th$#1#2$}%
    \setbox\tw@\hbox{\m@th$#1%
       \widehat{%
          \vrule\@width\z@\@height\ht\z@
          \vrule\@height\z@\@width\wd\z@}$}%
    \dp\tw@-\ht\z@
    \@tempdima\ht\z@ \advance\@tempdima2\ht\tw@ \divide\@tempdima\thr@@
    \setbox\tw@\hbox{%
       \raise\@tempdima\hbox{\scalebox{1}[-1]{\lower\@tempdima\box\tw@}}}%
    {\ooalign{\box\tw@ \cr \box\z@}}}
\makeatother

\numberwithin{equation}{section}

\renewcommand{\u}{{\bf u}}
\newcommand{\f}{{\bf f}}
\newcommand{\x}{{\bf x}}
\newcommand{\X}{{\bf X}}
\newcommand{\real}{\operatorname{Re}}
\newcommand{\imag}{\operatorname{Im}}
\newcommand{\iim}{\mathrm{i}}

\newcounter{question}
\newenvironment{myquestion}{%
  \par\sffamily\small
  \stepcounter{question}
  \begin{center}
    \begin{minipage}{0.8\textwidth}%
      \noindent{\scshape Question~\arabic{question}:\ \ }\itshape}{%
    \end{minipage}
  \end{center}
}

\newcommand{\leavethisout}[1]{}

\newcommand{\colorhline}{\arrayrulecolor{lightgray}\hline}
\newcommand{\colorvline}{\color{lightgray}\vrule}

\newcommand{\myrevision}[1]{#1}
\newcommand{\myrevisiontwo}[1]{{\color{red}#1}}
\renewcommand{\myrevisiontwo}[1]{#1}

\begin{document}

\title{An immersed boundary model of the cochlea with parametric
  forcing} 

\author{
  William Ko\footnotemark[2]
  \and 
  John M. Stockie\footnotemark[2]
}

\maketitle

\renewcommand{\thefootnote}{\fnsymbol{footnote}}
\footnotetext[2]{Department of Mathematics, Simon Fraser University, 
  8888 University Drive, Burnaby, BC, V5A 1S6, Canada
  ({\tt wka11@sfu.ca, jstockie@sfu.ca}).}


\begin{abstract}
  \myrevision{The cochlea or inner ear has a remarkable ability to
    amplify sound signals.  This is understood to derive at least in
    part from some active process that magnifies vibrations of the
    basilar membrane (BM) and the cochlear partition in which it is
    embedded, to the extent that it overcomes the effect of viscous
    damping from the surrounding cochlear fluid.  Many authors have
    \myrevisiontwo{associated}
    this amplification ability to some type of mechanical
    resonance within the cochlea, however there is still no consensus
    regarding the precise 
    \myrevisiontwo{cause of amplification.}  
    Our work is inspired by experiments
    showing that the outer hair cells within the cochlear partition
    change their lengths when stimulated, which can in turn cause
    periodic distortions of the BM and other structures in the cochlea.
    This paper investigates a novel fluid-mechanical resonance mechanism
    that derives from hydrodynamic interactions between an oscillating
    BM and the surrounding cochlear fluid.}
  We present a model of the cochlea based on the immersed boundary
  method, in which a small-amplitude periodic internal forcing due to
  outer hair cells can induce parametric resonance.  A Floquet stability
  analysis of the linearized equations demonstrates the existence of
  resonant (unstable) solutions within the range of physical parameters
  corresponding to the human auditory system.  Numerical simulations of
  the immersed boundary equations support the analytical results and
  clearly demonstrate the existence of \myrevision{resonant solution
    modes.  These results are} then used to illustrate the influence of
  parametric resonance on wave propagation along the BM and explicit
  comparisons are drawn with results from another two-dimensional
  cochlea model.
  \end{abstract}



\pagestyle{myheadings}
\thispagestyle{plain}
\markboth{W. KO AND J.~M. STOCKIE}
	{COCHLEA MODEL WITH PARAMETRIC FORCING}

\section{Introduction}

\begin{figure}
  \centering
  \includegraphics[width=0.95\textwidth]{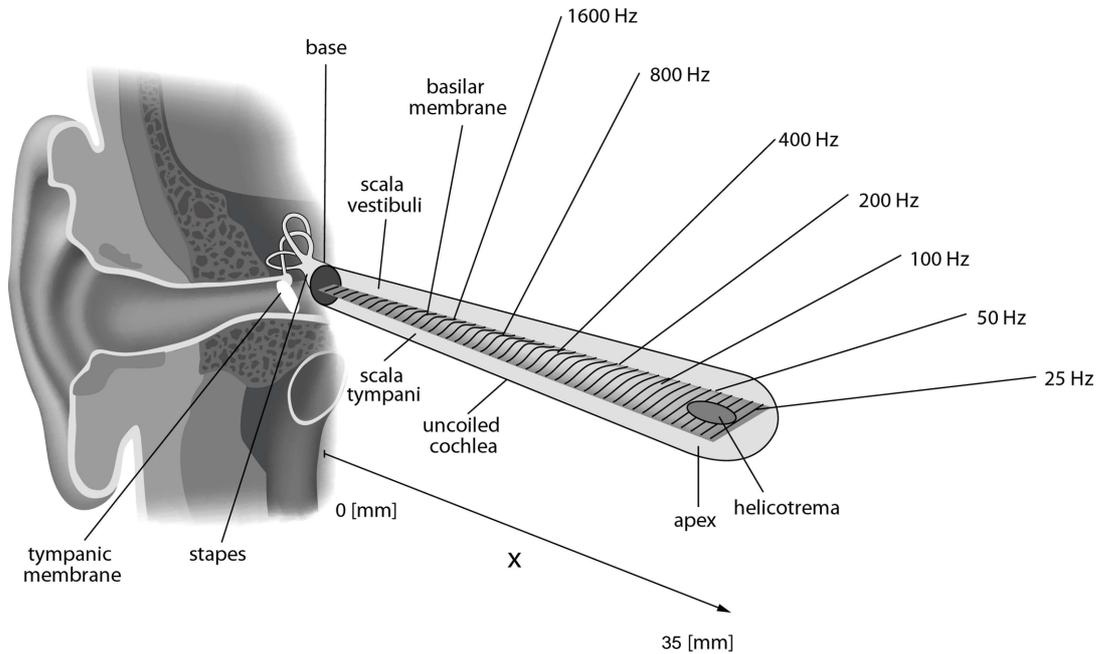}
  \caption{Diagram of an uncoiled cochlea and BM, showing the relative
    location of maximal BM response for several sound frequencies in the
    human audible range.  \myrevision{This is a simplified view of the
      cochlear partition, depicting only the BM and omitting other
      structural components such as the organ of Corti, outer hair
      cells, and tectorial membrane.}
    Source:~\cite[Fig.~2]{kern-etal-2008} (distributed under the
    Creative Commons License).}
  %
  \label{fig:wiki-cochlea}
\end{figure}

The cochlea or inner ear is a fluid-filled, spiral-wound cavity that is
\myrevision{the central source of frequency selectivity in the hearing
  system of} humans and many other animals.  \myrevision{The cochlea is
  divided on its primary axis into two main fluid chambers, the scala
  tympani and scala vestibuli, by a structure known as the cochlear
  partition (CP).  The CP itself consists of an elastic membrane known
  as the basilar membrane (BM), on top of which is mounted the organ of
  Corti containing the mechanosensitive hair cells that are the primary
  sensory receptors in the ear.  Figure~\ref{fig:wiki-cochlea} contains
  a picture of an unwound cochlea with a simplified view of the CP
  showing only the BM.}  Sound vibrations entering the outer ear are
transferred to the cochlear duct and BM by the stapes, and then
propagate from base to apex along the BM and the surrounding fluid.  The
BM has a fine frequency tuning ability that distinguishes
\myrevision{between different frequencies by localizing the peak
  amplitude of incident travelling waves.}  This ``place theory'' was
developed and validated experimentally by
von~B\'{e}k\'{e}sy~\cite{vonbekesy-1960} and shows that the peak
location is closer to the base for high frequencies and to the apex for
low frequencies, as illustrated in Figure~\ref{fig:wiki-cochlea}.

A second defining feature of the cochlea, which sets it apart from other
sensory receptors, is its ability to amplify extremely weak sound
signals that would otherwise be immediately damped out due to viscous
dissipation in the cochlear fluid.  Many authors have attributed this
amplification ability to some active process related to resonance, which
experiments have connected with mechanical properties of various
structures making up the \myrevision{CP}~\cite{dallos-1992,
  deboer-nuttall-2010, hudspeth-1997, hudspeth-2008,
  nobili-mammano-ashmore-1998}.  
  \myrevision{In particular, the outer
  hair cells in the organ of Corti are stimulated by BM deflections
  caused by pressure waves travelling through the cochlear fluid.  
     \myrevisiontwo{The hair cell stimulation leads to 
     either \emph{somatic motility},
    wherein the hair cell changes its length in response to electrical
    signals induced when the hair bundle on its tip is deflected; or
    \emph{active hair bundle motility} in which the hair bundle itself
    generates additional forces that initiate a shearing action between
    the BM and the overlying tectorial membrane. }
  Both of these effects
  are believed to contribute to the cochlear active process, but there
  is still no consensus in the literature regarding the precise 
  \myrevisiontwo{cause of amplification}~\cite{ashmore-etal-2010,reichenbach-hudspeth-2014}.
  }

Many mathematical and computational models of the cochlea have appeared
since the seminal work of von~B\'{e}k\'{e}sy~\cite{vonbekesy-1960}.  The
earliest two-dimensional models of the cochlea describe the BM as a
collection of damped mass-spring systems and also reduce the fluid
dynamics to a simple linear potential flow~\cite{allen-1977,
  lesser-berkley-1972, neely-1981}.  The spring constant decreases
exponentially along the BM from base to apex, which coincides with BM
stiffness values measured in experiments~\cite{vonbekesy-1960}.  The BM
is treated as a passive structure to which is applied a sinusoidal
external forcing term that mimics the input of sound energy from the
stapes.  These models give predictions of BM dynamics that agree
qualitatively with the behaviours observed by von~B\'{e}k\'{e}sy.
Inselberg and Chadwick~\cite{inselberg-chadwick-1976} proposed a similar
model in which the BM is represented as an Euler-Bernoulli beam, and
showed not only that the place principle still holds but also that fluid
viscosity is required to obtain travelling wave solutions along the BM
as opposed to simply standing
waves~\cite{chadwick-inselberg-johnson-1976}.
Pozrikidis~\cite{pozrikidis-2008} revisited this last approach by
replacing the sinusoidal stapes motion with a point source at the stapes
position and a point sink at the round window, and then solving the
resulting equations using a boundary integral method.
\myrevision{Another noteworthy class of models based on transmission
  line equations was introduced in the pioneering work of
  Zweig~\cite{zweig-1991} and de~Boer~\cite{deboer-1995b} and has since
  been applied in many more recent studies such 
  as~\cite{epp-verhey-mauermann-2010, verhulst-dau-shera-2012}.}

To obtain a more realistic model of the fluid dynamics in the cochlea as
well as the hydrodynamic interactions between the fluid and BM, several
authors have exploited the immersed boundary (or IB) method.  The IB
method was originally developed by Peskin to simulate blood flow in the
beating heart~\cite{peskin-1977} and has since been applied to many
problems in biofluid dynamics, including the cochlea.  LeVeque et
al.~\cite{leveque-peskin-lax-1985, leveque-peskin-lax-1988} employed an
IB model in which the fluid obeys the linearized Stokes equations and
the elasticity of the cochlea is treated using simple elastic springs.
They derived an asymptotic solution for travelling waves along the BM,
from which they drew conclusions regarding the effects of fluid
viscosity on these waves.  Beyer and LeVeque~\cite{beyer-1992} performed
numerical simulations of a related IB model, with the primary difference
being that their fluid obeys the full (nonlinear) Navier-Stokes
equations.  All of the aforementioned immersed boundary models
approximate the geometry of the cochlea and BM by a straight (uncoiled)
configuration.  Although the curvature of the cochlear duct has a
relatively small influence on the fluid dynamics~\cite{geisler-1998},
there is nonetheless some evidence to suggest that within the most
tightly coiled apical region of the BM that is stimulated by the lowest
frequency sounds, curvature cannot be
ignored~\cite{manoussaki-dimitriadis-chadwick-2006}.  To this end, a
much more detailed IB model capturing the full 3D geometry of the
cochlea was developed in~\cite{givelberg-bunn-2003} that reproduced
important features of BM dynamics.


The goal of this paper is to use the IB method to investigate possible
\myrevision{resonant phenomena that contribute to mechanical
  amplification of basilar membrane oscillations in the cochlea.}  Our
study is inspired by three main observations:
\myrevision{first, direct experimental evidence that outer hair cells
  embedded within the organ of Corti in the CP undergo periodic
  contractions when 
  \myrevisiontwo{the ear is stimulated} 
  by sound waves}~\cite{hudspeth-2008,
  karavitaki-mountain-2007, martin-hudspeth-1999}; second,
\myrevision{the suggestion by several authors that
  these hair cell contractions can in turn modulate the 
  \myrevisiontwo{stiffness} in the
  BM~\cite{mammano-ashmore-1993, markin-hudspeth-1995,
    nilsen-russell-1999} which we take as an assumption;} and third, the
analytical results of Cortez et al.~\cite{cortez-etal-2004} that
uncovered parametric instabilities arising from the fluid-structure
interaction in internally-forced immersed boundaries with a time-varying
stiffness parameter.  Taken together, these three
\myrevision{observations} suggest that there is merit in investigating
the hypothesis that internal forcing in the basilar membrane due to
sound stimulation can engender parametric resonance in an IB model of
the cochlea.  

\myrevision{One of the main contributions of this work is to extend the
  previous parametric resonance analysis for an elastic membrane with a
  periodically-varying stiffness parameter~\cite{cortez-etal-2004}
  to the case where stiffness depends on both time and space.}
\myrevision{Our results may also contribute new insights into the
  understanding of fine-frequency tuning and amplification in the
  cochlea since many previous cochlea models (described above) treat the
  outer hair cell contractions as an external forcing, over-simplify the
  fluid mechanics, and/or ignore the fluid-structure interaction.  For
  example, it is common for papers to question the ability of linked
  mechanical oscillator models to capture cochlear amplification by
  arguing that damping due to viscosity of the cochlear fluid is simply
  too large~\cite{hudspeth-1997}.  This argument is valid if the active
  process driving the cochlea appears as an external forcing in the
  model, but not when the force acts internally through a parameter --
  in the latter case, parametric resonance can occur in which large
  amplitude oscillations (unbounded in the idealized linear case) arise
  even in the presence of damping~\cite{champneys-2009}.}
\myrevision{Finally, we draw a distinction between our approach and
  other models incorporating resonance effects that act \emph{locally},
  such as the transmission line models based on coupled mechanical
  oscillators where each line element resonates individually.  In
  contrast, our model exhibits \emph{global resonance} owing to the
  non-local nature of the coupling between the BM and the surrounding
  fluid.}

\section{Mathematical Model}
\label{sec:model}

\begin{figure}[bthp]
  \centering
  \small
  \begin{minipage}{0.15\textwidth}
    \begin{pspicture}(1,1)
      \psline{->}(0,1)
      \psline{->}(1,0)
      \put(-0.05,1.10){$y$}
      \put(1.05,-0.05){$x$}
    \end{pspicture}
  \end{minipage}
  \begin{minipage}{0.06\textwidth}
    \footnotesize
    \begin{center}
      $(+\infty)$\\
      $\uparrow$\\[1.7cm]
      0\\[1.4cm]
      $\downarrow$\\
      $(-\infty)$\\[0.3cm]
      \mbox{\ }
    \end{center}
  \end{minipage}
  \begin{minipage}{0.71\textwidth}
    \includegraphics[width=\textwidth]{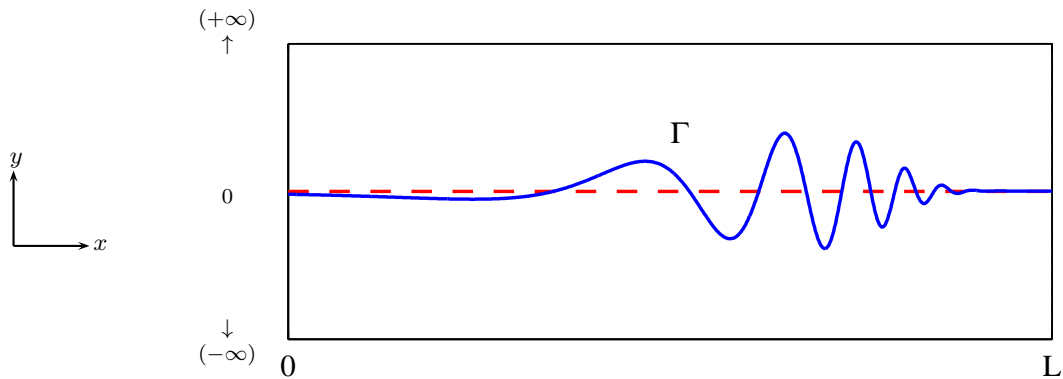}
  \end{minipage}
  \caption{Geometry of the 2D immersed boundary model for the cochlea.
    The deformed BM is represented by a solid blue line $\Gamma$, and
    the flat equilibrium state by a red dotted line.  
    \myrevision{To simplify the analysis, even symmetry is imposed across $x=0$ 
    and $x=L$ so that the solution is defined for all $x \in \mathbb{R}$,}
    although only values of $x\in[0,L]$ have physical relevance.}
  \label{fig:setup}
\end{figure}

Following the IB cochlea model derived by Peskin in~\cite{peskin-1976}
and developed in more detail in~\cite{leveque-peskin-lax-1988}, we
consider a simple 2D geometry pictured in Figure~\ref{fig:setup} in
which the cochlear duct of length $L$ is treated as a rectangular strip
$\Omega = \left[0,L\right]\times\mathbb{R}$, along the center of which
lies the BM.
\myrevision{This work is primarily concerned with the effects of
  parametric forcing on BM oscillations and so we simplify the model by
  isolating the membrane from any boundary effects due to the cochlear
  walls.  For the purposes of the mathematical analysis, we take the
  depth of the cochlear duct to be infinite ($y\to\pm\infty$) 
  \myrevisiontwo{which is
  consistent with the \emph{short-wave approximation}~\cite{shera-tubis-talmadge-2005} 
  and has been used previously by various authors}~\cite{keller-neu-1985,
    leveque-peskin-lax-1985}.  
    This assumption neglects effects such as
  coherent backscattering localized near the travelling wave 
  peak~\cite{shera-tubis-talmadge-2005}, but we have found that this does not
  significantly impact on the results of our stability analysis.
  Furthermore, we impose Neumann boundary conditions in the
  $x$-direction which is an assumption that has also been justified for
  other cochlea models~\cite{neely-1981, xin-2004}.}

The fluid in which the BM is immersed is governed by the incompressible
Navier-Stokes equations
\begin{align}
  \label{eq:dim-nse}
  \rho\left(\frac{\partial \u}{\partial t} + \u\cdot\nabla\u \right)
  &= -\nabla p + \mu\Delta\u + \f, 
  \\
  \label{eq:dim-incompressible}
  \nabla\cdot\u &= 0,
\end{align}
where
\myrevision{$\u(\x,t)$} is velocity, $p(\x,t)$ is pressure, $\rho$ is
density, and $\mu$ is viscosity.  The elastic force exerted on the fluid
by the membrane is given by
\begin{gather}
  \label{eq:force}
  \f(\x,t) = \int_0^L K(s,t)\, (\X_0-\X)\, \delta(\x-\X) \; ds,
\end{gather}
where $\delta(\x)$ is the two-dimensional Dirac delta function,
$\X(s,t)$ is the position of the membrane parameterized by the
Lagrangian coordinate $s \in [0,L]$, and $\X_0(s)=(s,0)$ is the
horizontal equilibrium or rest state.  This force can be thought of as
arising from a membrane that is connected to its resting position
via a series of Hookean springs with ``spring constant'' or stiffness
$K(s,t)$.  The elastic stiffness parameter is a function of
time and space given by
\begin{gather}
  \label{eq:Kst}
  K(s,t) = \sigma e^{-\lambda s}\big(1+2\tau\sin(\omega t)\big), 
\end{gather}
where $\sigma$ is the time-averaged elastic stiffness constant
\myrevision{(units of $\text{g}\,\text{cm}^{-2}\,\text{s}^{-2}$)}\ and
$\lambda$ captures the spatial variation in stiffness along the BM.  The
value of $\lambda \approx 1.4\, \text{cm}^{-1}$ has been determined for
a human cochlea experimentally by
von~B\'{e}k\'{e}sy~\cite{vonbekesy-1960} and others, based on the
observation that the BM stiffness at the apex is approximately two
orders of magnitude smaller than that at the base, and that it decays
roughly exponentially.  The time-dependent factor in the stiffness
encapsulates the parametric forcing with amplitude $\tau$ and frequency
$\omega$ arising from outer hair cells that contract/expand in response
to BM oscillations~\cite{brownell-etal-1985, howard-hudspeth-1988,
  hudspeth-2008}.  Note in particular that the forcing frequency
$\omega$ is taken to be the same as that of the input sound signal
\myrevision{and is also constant in space, which assumes that the outer
  hair cells contract in synchrony along the entire CP.  This is in
  contrast with most other models of the cochlea that consider hair cell
  contractions in response to local stimuli, which would correspond to a
  stiffness parameter having spatiotemporal dependence that is not
  separable.  There is nothing in our analytical approach that prevents
  us from considering this situation, but we nonetheless restrict our
  attention to the separable form in \eqref{eq:Kst} for the sake of
  simplicity.}

We close the system of equations with an expression for the membrane
motion
\begin{gather}
  \label{eq:dXdt}
  \frac{\partial \X}{\partial t} 
  = \u(\X,t) 
  = \int_\Omega \u(\x,t)\, \delta(\x-\X) \; d\x ,
\end{gather}
which is equivalent to the fluid no-slip condition along the immersed
boundary.

To simplify the analysis in the remainder of this paper, we
non-dimensionalize the problem by introducing the following scalings
\begin{gather}
  \x = \frac{L\widetilde{\x}}{\pi} , \quad 
  t  = \frac{\widetilde{t}}{\omega} , \quad
  \u = U_c\widetilde{\u}, \quad 
  p  = P_c \widetilde{p}, \quad
  \X = \frac{L\widetilde{\X}}{\pi} , \quad
  s  = \frac{L\widetilde{s}}{\pi} ,
\end{gather}
where the tildes denote a non-dimensional quantity.  The characteristic
velocity and pressure scales, $U_c$ and $P_c$, will be specified
shortly. The horizontal extent of the rescaled domain $\widetilde{\Omega} =
[0,\pi]\times\mathbb{R}$ is chosen for reasons of mathematical
convenience, in order to eliminate a factor of $\pi$ that would
otherwise appear in the solutions derived in section~\ref{sec:analysis}.

Substituting the above variables into the governing
equations~\eqref{eq:dim-nse}--\eqref{eq:dXdt} yields
\begin{align}
  \frac{\partial\widetilde{\u}}{\partial\widetilde{t}} +
  \widetilde{\u}\cdot\widetilde{\nabla} \widetilde{\u}
  &= -\widetilde{\nabla} \widetilde{p} + \nu\widetilde{\Delta}
  \widetilde{\u} + \widetilde{\f}, \\
  \widetilde{\nabla}\cdot\widetilde{\u} &= 0,\\
  \widetilde{\f}(\,\widetilde{\x},\widetilde{t}\,) &=
  \int_0^\pi \widetilde{K}(\widetilde{\X}_0-\widetilde{\X})\, 
  \widetilde{\delta}(\widetilde{\x}-\widetilde{\X}) \; d\widetilde{s},\\ 
  \widetilde{K}(\,\widetilde{s},\widetilde{t}\,) &= \kappa e^{- \alpha
    \widetilde{s}}(1+2\tau\sin \widetilde{t}\,),	\\ 
  \frac{\partial\widetilde{\X}}{\partial\widetilde{t}} &=
  \widetilde{\u}(\,\widetilde{\X},\widetilde{t}\,). 
\end{align}
The velocity and pressure scales are chosen as 
\begin{gather}
  U_c = \frac{L\omega}{\pi} \qquad \text{and} \qquad
  P_c = \frac{\rho\omega^2L^2}{\pi^2}, \\
  \intertext{to reduce the number of dimensionless parameters
    appearing in the equations to four, consisting of $\tau$ plus the
    three new quantities}%
  \nu    = \frac{\mu\pi^2}{\rho L^2 \omega}, \qquad
  \kappa = \frac{\sigma\pi}{\rho L \omega^2}, \qquad 
  \alpha = \frac{\lambda L}{\pi}.
\end{gather}
From this point onwards, the tildes will be dropped.

We next proceed to linearize the governing equations for the purpose of
making the analysis tractable.  The typical vertical displacement of the
BM is approximately $10^{-6}$~cm which is six orders of magnitude
smaller than its length of 3.5~cm~\cite{vonbekesy-1960}.  This implies
that the flow Reynolds number is very low, on the order of $10^{-6}$ or
less, and so nonlinear effects can be ignored.
Assuming that there is negligible coupling \myrevision{in the membrane}
along the longitudinal ($x$) direction, we only consider membrane
displacements in the $y$-direction.  Another major simplification is
achieved by eliminating the delta functions and reformulating the
problem in terms of jump conditions across the BM.  To do so, we
integrate the governing equations across the membrane at its linearized
rest state $y=0$~\cite{lai-li-2001}, yielding the alternate system of
equations
\begin{align}
  \label{eq:nondim-momentum}
  \frac{\partial \u}{\partial t}
  &= - \nabla  p + \nu \Delta\u, \\
  \label{eq:nondim-incompressible}
  \nabla\cdot\u &= 0,
\end{align}
away from the BM and
\begin{align}
  \label{eq:pressure-jump}
  \llbracket p \rrbracket &= -\kappa e^{-\alpha x}(1+2\tau\sin t)\, h(x,t),\\
  \label{eq:bc-u}
  u(x,0,t) &= 0,	\\
  \label{eq:bc-v}
  v( x,0, t) &= \frac{\partial  h}{\partial t},
\end{align}
across the membrane where 
\myrevision{$h(x,t)$ represents the vertical membrane displacement,
  $\llbracket p \rrbracket (x,t) = p(x,0^+,t)- p(x,0^-,t)$ is the jump
  in pressure across the membrane, and $u(x,y,t)$ and $v(x,y,t)$ are the
  horizontal and vertical components of the vector velocity $\u$.}
Eqs.~\eqref{eq:nondim-momentum}--\eqref{eq:bc-v} were studied
analytically in~\cite{leveque-peskin-lax-1988} without the time-varying
stiffness parameter (that is, with $\tau=0$).

\section{Floquet Analysis}
\label{sec:analysis}

Owing to the presence of a time-varying parameter in the system, we
invoke Floquet theory~\cite{champneys-2009} to analyze the
stability of perturbations of the membrane from its resting state.
Floquet theory assumes a solution of the form
\begin{gather}
  \label{eq:solution-form}
  u(\x,t) = e^{\gamma t}\, P(\x,t),
\end{gather}
where the function $P(\x,t)$ is periodic in time with period $2\pi$.
The Floquet exponent $\gamma \in \mathbb{C}$ determines the stability of
the solution as $t\to\infty$.  For the purposes of our analysis, we
extend the $x$-domain to $[-\pi,\pi]$ and impose even symmetry across
$x=0$.  We then take the unknown variables to be of the form
\begin{align}
  \label{eq:series-u}
  u(x,y,t) &= e^{\gamma t} \sum_{n = -\infty}^\infty \sum_{k=-\infty}^\infty
  u^n_k(y)\, e^{\iim nt} e^{\iim kx}, \\
  \label{eq:series-v}
  v(x,y,t) &= e^{\gamma t} \sum_{n = -\infty}^\infty \sum_{k=-\infty}^\infty
  v^n_k(y)\, e^{\iim nt} e^{\iim kx}, \\
  \label{eq:series-p}
  p(x,y,t) &= e^{\gamma t} \sum_{n = -\infty}^\infty \sum_{k=-\infty}^\infty
  p^n_k(y)\, e^{\iim nt} e^{\iim kx}, \\
  \label{eq:series-h}
  h(x,t) &= e^{\gamma t} \sum_{n = -\infty}^\infty \sum_{k=-\infty}^\infty
  h^n_k\, e^{\iim nt} e^{\iim kx},
\end{align}
where we have assumed that $P(\x,t)$ can be expanded using a Fourier
series in both space and time.  Our solution approach is similar to that
used by Cortez et al.~\cite{cortez-etal-2004} (modulo the corrections
in~\cite{ko-stockie-2012}) who analyzed the stability of a 2D circular
membrane in response to a perturbation in the form of a single Fourier
mode.  In contrast, here we must represent each solution mode as an
infinite Fourier series because of the mode-coupling through the spatial
non-uniformity in the stiffness parameter \myrevision{which will be shown
later in this section}.

We begin by finding solutions for the $y$-dependent Fourier coefficients
$u^n_k$, $v^n_k$ and $p^n_k$.  Take the divergence of the momentum
equations \eqref{eq:nondim-momentum} and apply the incompressibility
condition \eqref{eq:nondim-incompressible} to arrive at a Poisson
problem for pressure
\begin{gather}
  \Delta p 
  = \sum_{n,k = -\infty}^\infty 
  	\big( {-k^2} p^n_k(y) + {p^n_k}^{\prime\prime}(y)\big)\mathcal{E}^n_k 
  = 0.  
\end{gather}
Here we have simplified notation by setting $\mathcal{E}^n_k(x,t) =
\myrevision{\exp [{(\gamma+\iim n)t + \iim kx}]}$ and denoting
$y$-der\-iv\-a\-tives using primes.  The $\mathcal{E}^n_k$ are all
linearly independent and so we have a decoupled system of ordinary
differential equations for the $p^n_k(y)$
\begin{gather}
  -k^2 p^n_k(y)+  {p^n_k}^{\prime\prime}(y) = 0\quad \forall
  n,k\in\mathbb{Z},
\end{gather}
which after imposing boundedness in $y$ yields the pressure solution
\begin{gather}
  p^n_k(y) = \left\{
    \begin{aligned}
      &a^n_k e^{ ky} & \quad\text{if } y < 0, \\
      &b^n_k e^{-ky} & \quad\text{if } y > 0, \\
    \end{aligned}\right.
\end{gather}
for each $n,k\in \mathbb{Z}$, with constants $a^n_k$ and $b^n_k$ yet to
be determined.  Note that taking $k=0$ is valid in the above expression
since this implies simply a constant pressure in each sub-domain.  

We can now solve for the Fourier coefficients of the vertical velocity
component $v$ by substituting the series \eqref{eq:series-v} into the
momentum equation to obtain
\begin{gather}
  \sum_{n,k = -\infty}^\infty(\gamma + \iim n){v^n_k}(y) \mathcal{E}^n_k = 
  \sum_{n,k = -\infty}^\infty \big( {-p^n_k}^\prime(y) -
    \nu k^2{v^n_k}(y) + \nu {v^n_k}^{\prime\prime}(y) \big) \mathcal{E}^n_k. 
\end{gather}
This is equivalent to the infinite linear system of ordinary
differential equations
\begin{gather}
  {v^n_k}^{\prime\prime}(y) - (\beta^n_k)^2{v^n_k}(y) =
  \frac{1}{\nu}{p^n_k}^\prime(y)\quad \forall n,k\in\mathbb{Z} ,
\end{gather}
where
\begin{gather}
	\label{eq:beta}
  \beta^n_k = \sqrt{\frac{\gamma + \iim n}{\nu} + k^2}\quad \text{with}
  \; \real\{\beta^n_k\} > 0. 
\end{gather}
Assuming that $\gamma + \iim n \ne 0$ and $k \ne 0$, the solution is
\begin{gather}
  {v^n_k}(y) = \frac{1}{2\nu\beta^n_k} \left\{
    \begin{aligned}
      -\frac{2k\beta^n_k}{({\beta^n_k})^2-k^2}a^n_k\, e^{ky}
      &+ \left(\frac{k}{\beta^n_k-k}a^n_k +
\frac{k}{\beta^n_k+k}b^n_k\right)e^{\beta^n_k y}, 
      & \text{if } y < 0, \\	
      \frac{2\beta^n_k}{({\beta^n_k})^2-k^2}b^n_k\, e^{-ky}
      &- \left(\frac{k}{\beta^n_k-k}b^n_k +
\frac{k}{\beta^n_k+k}a^n_k\right)e^{-\beta^n_k y}, 
      & \text{if } y > 0.
    \end{aligned}
  \right.
\end{gather}
The continuity equation reduces to
\begin{gather}
  \iim ku^n_k(y) + {v^n_k}^\prime(y)  = 0 \quad \forall n,k\in\mathbb{Z},
\end{gather}
from which $u^n_k(y)$ are found to be
\begin{gather}
  {u^n_k}(y) = \frac{-\iim}{2k\nu} \left\{
    \begin{aligned}
      -\frac{2k^2}{({\beta^n_k})^2-k^2}a^n_k\, e^{ky}
      & + \left(\frac{k}{\beta^n_k-k}a^n_k +
\frac{k}{\beta^n_k+k}b^n_k\right)e^{\beta^n_k y}, 
      & \text{if } y < 0, \\
      -\frac{2k}{({\beta^n_k})^2-k^2}b^n_k\, e^{-ky}
      & + \left(\frac{k}{\beta^n_k-k}b^n_k +
\frac{k}{\beta^n_k+k}a^n_k\right)e^{-\beta^n_k y}, 
      & \text{if } y > 0.
    \end{aligned}
  \right.
\end{gather}	
We then impose the continuity conditions on velocity across the membrane
\begin{align}
  u^n_k(0^+) &= u^n_k(0^-) 
  = \frac{\iim}{2\nu}\left(\frac{1}{\beta^n_k+k}a^n_k +
    \frac{1}{\beta^n_k+k}b^n_k\right) 
  = 0,	\\
  v^n_k(0^+) &= v^n_k(0^-)
  = \frac{1}{2\nu\beta^n_k}\left( \frac{-k}{\beta^n_k+k}a^n_k +
    \frac{k}{\beta^n_k+k}b^n_k\right) 
  = (\gamma + \iim n) h^n_k,
\end{align}
which can be solved in terms of the $h^n_k$ as
\begin{align}
  a^n_k &= -\nu(\gamma+\iim n)\, \frac{\beta^n_k + k}{k}\,\beta^n_k h^n_k, \\
  b^n_k &= \phantom{-}\nu(\gamma+\iim n)\, \frac{\beta^n_k + k}{k}\,\beta^n_k h^n_k.
\end{align}
We note that the special cases $\gamma + \iim n = 0$ and $k=0$ both yield 
the trivial solution, $a^n_k=b^n_k=0$.

Using the jump condition \eqref{eq:pressure-jump}, we can now formulate
an infinite system of linear equations that connects all of the membrane
coefficients, $h^n_k$.  Substituting the solution for pressure into
\eqref{eq:pressure-jump}, we obtain
\begin{align}
  \begin{aligned}
    \sum_{n,k=-\infty}^\infty 2\nu(\gamma + \iim n)\, \frac{\beta^n_k
      + k}{k}\, \beta^n_k h^n_k \mathcal{E}^n_k 
    =
    \sum_{n,k=-\infty}^\infty -\kappa e^{-\alpha x}
    \left(1+2\tau\sin t\right) h^n_k \mathcal{E}^n_k, 
  \end{aligned}
\end{align}
provided that $\gamma + \iim n\ne0$ and $k\ne0$.  In the special cases
when $\gamma + \iim n = 0$ or $k=0$, these equations reduce to
\begin{align}
  0 = \sum_{n,k=-\infty}^\infty -\kappa e^{-\alpha x} (1+2\tau\sin t)
  h^n_k \mathcal{E}^n_k. 
\end{align}

In order to proceed any further, we first need to expand the exponential
($e^{-\alpha x}$) and sinusoidal ($\sin t$) terms in their respective
Fourier series.  For the time-dependent factor, we can write
$1+2\tau\sin t = 1 -\iim\tau e^{\iim t} + \iim \tau e^{-\iim t}$, but
the Fourier series for the exponential function does not converge
uniformly because $e^{-\alpha x}$ is not a periodic function on
$[0,\pi]$.  It is for this reason that we have extended the spatial
domain to $[-\pi,\pi]$, and then we simply need to remember that is only
the portion with $x\ge 0$ that is of physical interest. We instead use
the even periodic extension $e^{-\alpha|x|}$ of the elastic stiffness
function on this extended interval. Then, for $\gamma + \iim n\ne 0$ and
$k\ne 0$ we have
\begin{gather}
  \begin{aligned}
    \sum_{n,k=-\infty}^\infty &2\nu(\gamma + \iim n)
    \, \frac{\beta^n_k + k}{k}\, \beta^n_k h^n_k \mathcal{E}^n_k \\ 
    &= \sum_{n,k=-\infty}^\infty -\kappa 
    \left(\sum_{j=-\infty}^\infty c_je^{\iim jx}\right) 
    (1 - \iim\tau e^{\iim t} + \iim\tau e^{-\iim t}) h^n_k \mathcal{E}^n_k,
  \end{aligned}
  \label{eq:modes-eq1}
\end{gather}
while for $\gamma + \iim n = 0$ or $k=0$,
\begin{gather}
  0 = \sum_{n,k=-\infty}^\infty -\kappa 
  \left(\sum_{j=-\infty}^\infty c_je^{\iim jx}\right) 
  (1 - \iim\tau e^{\iim t} + \iim\tau e^{-\iim t}) h^n_k \mathcal{E}^n_k,
  \label{eq:modes-eq2}
\end{gather}
where 
\begin{gather}
  c_j = \alpha\frac{1-(-1)^je^{-\alpha\pi}}{\pi(\alpha^2+j^2)}
\end{gather}
are the Fourier coefficients of $e^{-\alpha|x|}$ on $[-\pi,\pi]$.  The
complex exponential terms involving $x$ and $t$ in
Eqs.~\eqref{eq:modes-eq1} and~\eqref{eq:modes-eq2} introduce a shift in
the indices of $\mathcal{E}^n_k$ in both $n$ and $k$, which has the
effect of coupling the corresponding modes.  We can then rearrange 
the sum in order to gather together all terms involving the common
expression $\mathcal{E}^n_k$, and hence obtain
\begin{gather}
  \label{eq:system1}
  \begin{aligned}
    \frac{2\nu^2}{\kappa}(\beta^n_k - k) \, (\beta^n_k + k)^2 \,
    \frac{\beta^n_k}{k} h^n_k + \sum_{j=-\infty}^\infty c_{k-j} h^n_{j} 
    = \iim\tau \sum_{j=-\infty}^\infty c_{k-j} ( h^{n-1}_{j} - h^{n+1}_{j}),
  \end{aligned}
\end{gather}
for $\gamma + \iim n\ne0$ and $k\ne0$ and
\begin{gather}
  \label{eq:system2}
  \sum_{j=-\infty}^\infty c_{k-j} h^n_{j} 
  = \iim\tau \sum_{j=-\infty}^\infty c_{k-j} ( h^{n-1}_{j} - h^{n+1}_{j}),
\end{gather}
for either of the special cases $\gamma + \iim n = 0$ or $k=0$.  These
last two equations comprise an infinite linear system for the $h^n_j$ in
which the spatially-dependent stiffness introduces a simultaneous
coupling between all spatial modes that is not present in the
spatially-uniform ($\alpha=0$) case from~\cite{cortez-etal-2004}.

Because we are interested in investigating the stability of the
parametrically forced problem, and in particular finding the stability
boundary in parameter space, we only have to look for periodic solutions
of Eqs.~\eqref{eq:system1}--\eqref{eq:system2} in the situation when
$\real\{\gamma\} =0$.  Furthermore, there are only two distinct values of
$\gamma$ that are of actual interest for determining the stability
boundary: the first corresponds to $\gamma=0$ and will be referred to as
the {\em harmonic} case; and the second, $\gamma = \frac{1}{2}\iim$, is
called the {\em subharmonic} case.  To ensure that the solution
$h(x,t)$ is real-valued, we impose \emph{reality conditions} for the
Fourier coefficients that apply to both time and space indices.  In
general, the reality condition for a two-dimensional Fourier series is
$h^n_k = \bar{h}^{-n}_{-k}$ where the overbar denotes the complex
conjugate.  However, we have to consider the harmonic and subharmonic
cases separately when applying the condition in the temporal modes.
Furthermore, we want to ensure an even spatial symmetry in our solutions
which leads to a decoupling in the reality condition.  Consequently, the
reality conditions that we impose are
\begin{alignat}{3}
  h^n_{-k} &= h^n_k, \label{eq:reality1}\\ 
  h^{-n}_k &=
  \begin{cases}
    \bar{h}^n_k,      & \text{for the harmonic case ($\gamma=0$)},\\
    \bar{h}^{n-1}_k,  & \text{for the subharmonic case
($\gamma=\frac{1}{2}\iim$)}.
  \end{cases}
  \label{eq:reality2}
\end{alignat}
As a result, the reality conditions introduce a certain symmetry between
the Fourier coefficients such that we only need to consider non-negative
values of $n$ and $k$.

These choices of harmonic and subharmonic values for $\gamma$ can be
justified as follows. The solution form \eqref{eq:solution-form} implies
that 
\begin{gather}
  u(\x,t+2\pi n) = e^{\gamma(t+2\pi n)}P(\x,t) = \xi^n u(\x,t),
\end{gather}
for any positive integer $n$ where $\xi = e^{\gamma 2\pi}$ and $t$ is
fixed.  As $n\to\infty$ the long term behaviour of the solution depends
the value of $\xi$. We conclude that solutions are stable if $|\xi|<1$
and unstable if $|\xi|>1$, where the special values $\xi= \pm 1$
correspond to periodic solutions that define the marginal stability
boundaries separating stable and unstable solutions.  If $\gamma = 0$,
then $\xi = 1$ and
\begin{gather}
  u(\x,t+2\pi) = u(\x,t),
\end{gather}
which is the $2\pi$-periodic harmonic solution.
If $\gamma = \frac{1}{2}\iim$, then $\xi = -1$ and
\begin{gather}
  u(\x,t+2\pi) = -u(\x,t), \quad
  u(\x,t+4\pi) = u(\x,t),
\end{gather}
which is the period-doubling subharmonic solution.  These same
relationships hold for all of the dependent variables.

We are aware of no analytical method for determining solutions to the
infinite system~\eqref{eq:system1}--\eqref{eq:reality2} explicitly, and
so we truncate the system at values of $n = 0,1,\dots,N$ and $k = 1,2,
\dots,M$ and then approximate the solutions numerically.  For $k=0$, we
have that $h^n_0=0$ for each $n$ and so there is no need to include them
in the linear system.  The truncated system of equations can therefore
be represented as a matrix equation
\begin{gather}
  \label{eq:matrix}
  A \vec{h} = \tau B \vec{h},
\end{gather}
where
\begin{gather}
  \vec{h} = 
  [\, \ldots, \;\real(h^n_k), \;\imag(h^n_k), \; \real(h^n_{k+1}), \;
  \imag(h^n_{k+1}), \; \ldots\, ]^T
  \label{eq:hvector}
\end{gather}
is a vector of length $2M(N+1)$ containing all unknown coefficients, $A$
(and $B$) are block diagonal (tridiagonal) matrices respectively, both
of block dimension $(N+1)\times (N+1)$ where each block is size
$2M\times 2M$.  The block diagonal matrix $A$ can be expressed as $A =
\diag(A^0, A^1,\ldots,A^N)$ where each block has the form
\begin{align}
  A^n &= \left[
    \begin{matrix}
      C_{1,1}	+ D^n_1	& C_{1,2} 	& \hdots	& C_{1,M}	\\
      C_{2,1} 	& C_{2,2}+D^n_2 &		& C_{2,M}	\\
      \vdots 		& 	  	& \ddots	& \vdots	\\
      C_{M,1} 	& C_{M,2} 	& \hdots 	& C_{M,M}+D^n_M
    \end{matrix}
  \right],\\
  \intertext{with}
  C_{k,j} &= \left[
    \begin{matrix}
      c_{k-j} + c_{k+j} & 0 \\
      0	& c_{k-j} + c_{k+j}
    \end{matrix}
  \right],\\
  \label{eq:block-D}
  D^n_k &= \frac{2\nu^2}{\kappa k}\left[
    \begin{array}{rr}
      \real\{(\beta^n_k-k)(\beta^n_k+k)^2\beta^n_k\} &
      -\imag\{(\beta^n_k-k)(\beta^n_k+k)^2\beta^n_k\} \\
      \imag\{(\beta^n_k-k)(\beta^n_k+k)^2\beta^n_k\} &
      \real\{(\beta^n_k-k)(\beta^n_k+k)^2\beta^n_k\}
    \end{array}
  \right].
\end{align}
For the harmonic case, $D^0_k$ is simply the $2\times 2$ zero matrix.
The block tridiagonal matrix $B$ has the form
\begin{gather}
  B = \left[
    \begin{matrix}
      \widecheck{B}& \widetilde{B} 	 	 	\\
      \widehat{B}	&  0  	&-\widehat{B}  	\\
      & \ddots&\ddots	& \ddots\\
      &	&\widehat{B}& 0	&-\widehat{B}\\
      &  	&	& \widehat{B}& 0
    \end{matrix}
  \right],
\end{gather}
where
\begin{gather}
  \label{eq:Bhat}
  \widehat{B} = \left[
    \begin{matrix}
      \widehat{C}_{1,1}	& \widehat{C}_{1,2} & \hdots	& \widehat{C}_{1,M}
      \\
      \widehat{C}_{2,1} & \widehat{C}_{2,2} &		& \widehat{C}_{2,M}
      \\
      \vdots 	& 	  & \ddots	& \vdots	\\
      \widehat{C}_{M,1} & \widehat{C}_{M,2} & \hdots 	& \widehat{C}_{M,M}
    \end{matrix}
  \right], \quad
  \widehat{C}_{k,j} = \left[
    \begin{matrix}
      0 	& -c_{k-j} - c_{k+j} \\
      c_{k-j} + c_{k+j}	& 0
    \end{matrix}
  \right].
\end{gather}
The reality conditions determine the first block row of $B$.  For
harmonic solutions, $\widecheck{B}$ is the $2M\times 2M$ zero matrix
while $\widetilde{B}$ has the same form as $\widehat{B}$
in~\eqref{eq:Bhat} except that the $2\times2$ sub-blocks are
\begin{gather}
  \widetilde{C}_{k,j} = \left[
    \begin{matrix}
      0 & 2c_{k-j} + 2c_{k+j} \\
      0	& 0
    \end{matrix}
  \right].
\end{gather}
For subharmonic solutions $\widetilde{B} = -\widehat{B}$, and
$\widecheck{B}$ consists of the $2\times2$ sub-blocks
\begin{gather}
  \widecheck{C}_{k,j} = \left[
    \begin{matrix}
      0 	& c_{k-j} + c_{k+j} \\
      c_{k-j} + c_{k+j}	& 0
    \end{matrix}
  \right].
\end{gather}

Now that the entries in the matrices $A$ and $B$ are all known, we can
view \eqref{eq:matrix} as an eigenvalue problem
\begin{gather}
  \label{eq:eig}
  A^{-1}B \vec{h} = \frac{1}{\tau} \vec{h},
\end{gather}
where the key determinant of stability is the eigenvalue
$\frac{1}{\tau}$.  In particular, we are only concerned with those
physically-relevant values of $\tau$ that are real and are less than
$\frac{1}{2}$, since this ensures that the elastic stiffness function
$K(s,t)$ is real and non-negative.

\section{Natural Modes for an Unforced Membrane}

Before solving the parametrically forced problem, we first examine the
stability of the unforced membrane corresponding to $\tau=0$.  Previous
stability analyses of the immersed boundary method showed that unforced
membranes are always stable~\cite{cortez-etal-2004, stockie-wetton-1995}
and we expect a similar result here.  The Fourier coefficients of the
unforced solution satisfy
\begin{gather}
  \label{eq:nat1}
  \frac{2\phi\gamma}{k\nu} \sqrt{\frac{\gamma}{\nu} + k^2}
  \left( k + \sqrt{\frac{\gamma}{\nu} + k^2}\, \right) h^0_k 
  + \sum_{j=-\infty}^\infty c_{k-j} h^0_{j} = 0
\end{gather}
when $k\ne0$, and
\begin{gather}
  \label{eq:nat2}
  \sum_{j=-\infty}^\infty c_{k-j} h^0_{j} = 0
\end{gather}
when $k=0$. We have introduced the new dimensionless parameter $\phi =
{\nu^2}/{\kappa} = {\pi^3\mu^2}/{(\rho\sigma L^3)}$, which is a measure
of the relative importance of viscous fluid force relative to elastic
membrane force.  The system of equations~\eqref{eq:nat1}
and~\eqref{eq:nat2} can be written simply as a matrix system $T\vec{h}^0
= {\bf 0}$, where the entries of the matrix $T$ depend on the parameters
$\phi$, $\gamma/\nu$ and $\alpha$.  This linear system has non-trivial
solutions only if $\det(T) = 0$, and so by fixing values of $\phi$ and
$\alpha$ we can determine values of $\gamma/\nu$ such that solutions to
the homogeneous system exist.  In practice, we proceed by first
truncating the infinite series in~\eqref{eq:nat1} and in~\eqref{eq:nat2}
to $M$ terms, and then computing the determinant numerically.

\begin{figure}[bthp]
  \centering
  \includegraphics{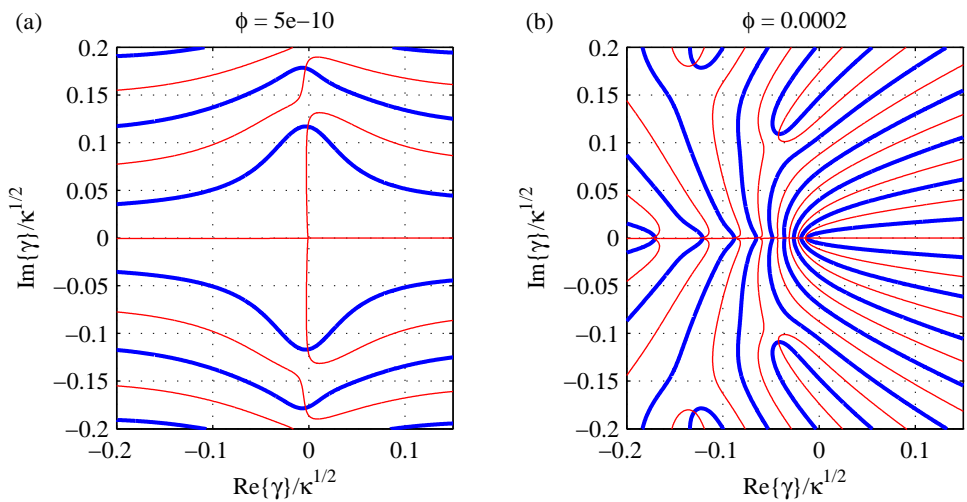}
  \caption{\myrevision{Zero contours of the real (thick, blue) and
      imaginary parts (thin, red) of $\det(T)$ for: (a) $\phi =
      5\times10^{-10}$ and (b) $\phi = 2\times10^{-4}$.  In both cases,
      we take $\alpha=1.56$ and use $M=20$ modes.  The resonant modes
      correspond to intersection points between the real and
      imaginary contours.}}
  \label{fig:nat}
\end{figure}

Figure~\ref{fig:nat} depicts the zero contours of the real and imaginary
parts of $\det(T)$ for the specific choice of parameters $\alpha=1.56$
and $M = 20$, and two values of $\phi = 5\times10^{-10},\;
2\times10^{-4}$.  The points where the contours intersect correspond to
the natural modes of the system.  
\myrevision{Observe that $\real\{\gamma\}$ is always negative
  \myrevisiontwo{at the intersection points}, from
  which we conclude that all solution modes are stable.}  We see next
how the behaviour of the natural modes depends on $\phi$, the relative
strength of fluid viscosity to membrane stiffness.  For the relatively
small value of $\phi = 5\times10^{-10}$,
\myrevision{the dominant modes (which are slowest to decay) have
  non-zero $\imag\{\gamma\}$ and are therefore oscillatory.}  When
$\phi$ is increased to $2\times10^{-4}$, viscosity has a much stronger
influence and the dominant (lowest wavenumber) modes decay without
oscillations, although decaying oscillatory solutions still do exist.
In both cases, the unforced modes always decay in time and hence any
periodic or unstable solutions must arise from a periodic modulation of
the membrane stiffness.

\section{Parametrically Forced Pure-Tone Response}

In this section, we demonstrate that an internally-forced membrane can
generate travelling wave solutions that are similar to solutions
obtained from other models that impose an external forcing.  We compute
harmonic solutions to \eqref{eq:system1}--\eqref{eq:system2}
numerically and then choose the smallest physically-allowable value of
$\tau$ along with its corresponding eigenvector.  We can reconstruct a
periodic solution for $h(x,t)$ using the series representation
\eqref{eq:series-h}, which is then compared to the travelling wave
solutions from~\cite{leveque-peskin-lax-1988} obtained when a pure-tone
external forcing is applied.  We take parameter values 
from~\cite{leveque-peskin-lax-1988} (listed in Table~\ref{tab:params}) that
are also consistent with parameters reported elsewhere in the literature
for the human cochlea, with the only exception being the elastic
stiffness value.  Other cochlea models take values of $\sigma$ that
range from $1\times10^7$~\cite{lesser-berkley-1972} to
$2\times10^9\,\text{g}\,\text{cm}^{-2}\text{s}^{-2}$~\cite{allen-1977};
however, we have chosen the smaller value of 
\myrevision{$\sigma=6\times 10^5\,\text{g}\,\text{cm}^{-2}\text{s}^{-2}$} for
the purposes of this paper in order to permit a direct comparison with
the results in~\cite{leveque-peskin-lax-1988} for a similar IB model.
\begin{table}
  \caption{Parameter values (or ranges) used in the cochlea
    simulations.  Values are taken from~\cite{leveque-peskin-lax-1988}
    and (with the exception of $\sigma$) correspond to the human
    cochlea.} 
  \centering\footnotesize
  \renewcommand{\extrarowheight}{0.1cm}
  \begin{tabular}{|cll|}
    \hline
    \multicolumn{3}{|l|}{\emph{Physical parameters (cgs units):}}\\
    & fluid density           & $\rho  = 1.0\, \text{g}\,\text{cm}^{-3}$\\
    & fluid viscosity         & $\mu   = 0.02\,\text{g}\,\text{cm}^{-1}\,\text{s}^{-1}$\\
    & elastic stiffness       & $\sigma = 6\times 10^5\,\text{g}\,\text{cm}^{-2}\,\text{s}^{-2}$\\
    & elastic stiffness decay rate & $\lambda = 1.4\, \text{cm}^{-1}$\\
    & basilar membrane length & $L = 3.5\, \text{cm}$\\
    & \myrevision{forcing} frequency         & $\omega \in [ 400, \; 1600 ]\;\text{s}^{-1}$\\
    & \myrevision{velocity scale} & \myrevision{$U_c = 
      \frac{\omega L}{\pi} \in [446, 1783]\, \text{cm}\, \text{s}^{-1}$}\\
    & \myrevision{pressure scale} & \myrevision{$P_c = 
      \frac{\rho \omega^2 L^2}{\pi^2} \in [2.0\times 10^5, 3.2 \times 10^6]\, \text{dyn}\,\text{cm}^{-2}$}\\
    \hline
    \multicolumn{3}{|l|}{\emph{Dimensionless parameters:}}\\
    & dimensionless decay rate & $\alpha=1.56$\\
    & \myrevision{forcing amplitude} & \myrevision{$\tau\in [0, 0.5]$} \\ 
    & dimensionless viscosity  & $\nu    \in [ 8.06\times 10^{-6},\; 1.61\times 10^{-4}]$\\
    & dimensionless stiffness  & $\kappa \in [ 0.135,\; 53.9 ]$\\
    & viscosity/stiffness ratio& \myrevision{$\phi=\frac{\nu^2}{\kappa}
    	= \frac{\pi^3\mu^2}{\rho\sigma L^3} = 4.8\times 10^{-10}$}\\
    \hline
  \end{tabular}
  \label{tab:params}
\end{table}

\begin{figure}[bthp]
  \centering
  \includegraphics{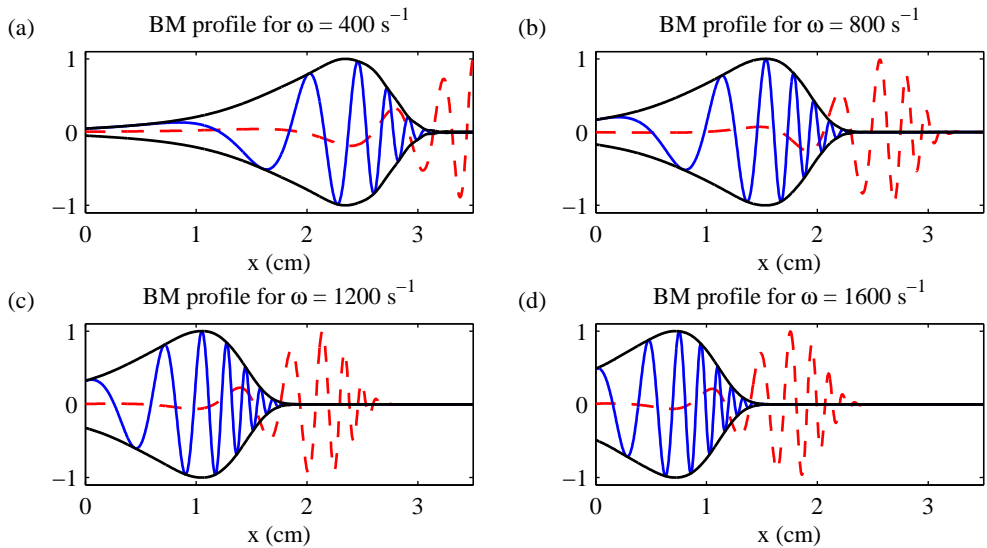}
  \caption{Normalized BM displacement profiles for 
    \myrevision{the harmonic (solid) and subharmonic (dashed) cases} at
    frequencies $\omega$ = 400, 800, 1200 and 1600 $\text{s}^{-1}$. }
  \label{fig:bm-profiles}
\end{figure}

Figure~\ref{fig:bm-profiles} shows solution profiles of the BM
displacement curve $h(x,t_{peak})$ for 
\myrevision{forcing frequencies} $\omega = 400$, $800$, $1200$ and
$1600\,\text{s}^{-1}$, where $t_{peak}$ represents the time when the
maximum vertical BM displacement occurs.  The wave envelope is
determined by computing the absolute value of a complex-valued function
whose real part is the BM profile and the imaginary part is its Hilbert
transform~\cite{bracewell-2000,kim-xin-2005}.  
The envelope is normalized so that the maximum height is 1.
\myrevision{ These wave envelopes have the characteristic asymmetric
  shape seen in experiments~\cite{reichenbach-hudspeth-2010}, exhibiting
  an amplitude that increases gradually from base to peak, followed by a
  sharp decline at the apical end.  The solid curve in each case
  corresponds to the harmonic mode for which the response frequency is
  equal to the forcing frequency.  A qualitatively similar result is
  obtained for subharmonic solutions (dashed curves) except that the
  response occurs at a frequency equal to half that of the internal
  forcing and so the wave profiles are shifted. This indicates that the
  location of the envelope peak depends on the response frequency and
  not the forcing frequency.}

%

\begin{figure}[bthp]
  \centering
  \includegraphics{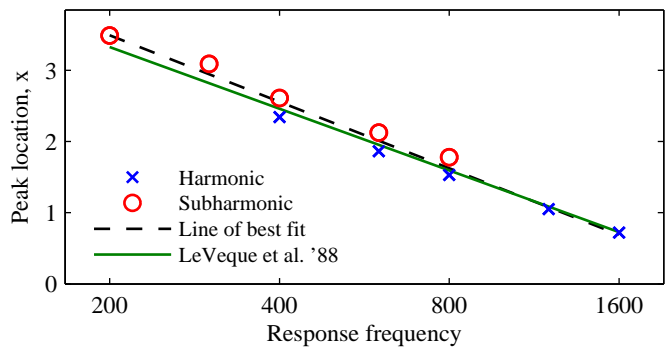}
  \caption{\myrevision{BM envelope peak location plotted against
      response frequency, with the horizontal axis (frequency) plotted
      on a log scale.  Harmonic $(\times)$ and subharmonic solutions
      $(\circ)$ have response frequencies $\omega$ and $\omega/2$,
      respectively.}}
  \label{fig:bm-peaks}
\end{figure}

\begin{figure}[bthp]
  \centering
  \includegraphics{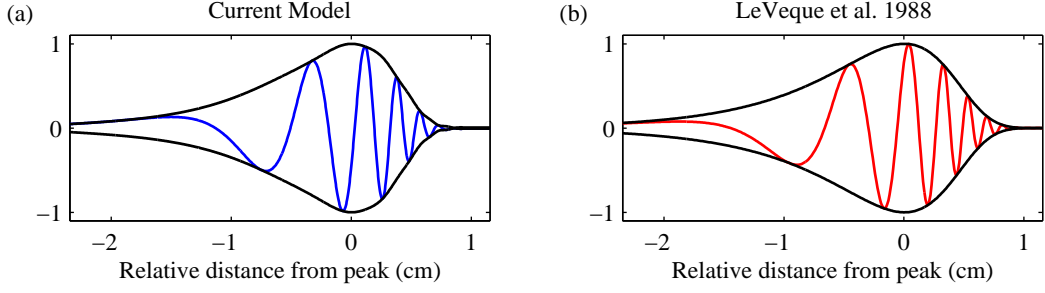}
  \caption{Normalized BM displacement profiles from our IB model
    (left) and model by LeVeque~et~al.~\cite{leveque-peskin-lax-1988} (right) 
    at a frequency of
    $\omega = 400\,\text{s}^{-1}$.}
  \label{fig:bm-compare}
\end{figure}

Through a combination of experiments and analysis,
\myrevision{von~B\'ek\'esy showed}~\cite{vonbekesy-1960} that the peak
location of the travelling wave envelope actually depends
logarithmically on the stimulus frequency.  Figure~\ref{fig:bm-peaks}
depicts the computed peak locations at various response frequencies, and our
results lie nearly on a straight line which is consistent with the
logarithmic dependence just mentioned.  This plot also includes the
asymptotic results derived from the model
in~\cite{leveque-peskin-lax-1988} as a solid line, which is clearly very
close to our own results.  As a further validation,
Figure~\ref{fig:bm-compare} compares our BM displacement curve with the
corresponding result from~\cite{leveque-peskin-lax-1988} for the
frequency $\omega=400\,\text{s}^{-1}$.  Both profiles are shifted so
that the wave envelope peak occurs at $x=0$, after which we can see that
the qualitative shape and solution envelope are quite similar.  These
numerical results demonstrate that external forcing is not required to
obtain a realistic travelling wave solution on the BM, and indeed that
parametric (internal) forcing in the BM stiffness can generate pure-tone
solutions consistent with that observed in other models.

\leavethisout{
  \begin{myquestion}
    If I were to suggest adding any figure, it would be something along
    the lines of Figs.~5 or~7 in S. Ramamoorthy, N.V. Deo, K. Grosh, ``A
    mechano-electro-acoustical model for the cochlea: Response to
    acoustic stimuli, \emph{J. Acoust. Soc. Amer.}
    \underline{121}(5):2758-2773, 2007.  By the way, their model appears
    to be one of the most well-described mechanical models, and so
    perhaps we can use this as the basis for our next publication (to
    PNAS?)  It would be helpful to go through~\cite{ashmore-etal-2010}
    (on which Grosh is a co-author) to figure out how Grosh's model
    figures into the grander scheme of cochlea modelling.
  \end{myquestion}
}

\section{Parametric Resonance}

\subsection{Case $\mathbf{\alpha=0}$}

To gain more insight into solutions of the eigenvalue problem
\eqref{eq:eig}, we begin by considering the simple case $\alpha=0$ where
the elastic stiffness does not depend on BM location and thus the
spatial Fourier modes are decoupled.  For each spatial wavenumber $k$ we
have
\begin{gather}
  \frac{2 \phi}{k}(\beta^n_k-k) (\beta^n_k + k)^2 \beta^n_k
  h^n_k + h^n_{k} = \iim\tau ( h^{n-1}_{k} - h^{n+1}_{k}) 
\end{gather}
for $n = 0,1,\ldots,N$.  This equation may be rewritten in matrix form
as 
\begin{gather}
  A_k \vec{h}_k = \tau B_k \vec{h}_k,
\end{gather}
where for each value of $k$ 
\begin{align}
  \vec{h}_k &= [ \, \real(h^0_k), \; \imag(h^0_k), \; \ldots, \;
  \real(h^{N}_k), \; \imag(h^{N}_k) \,]^T, \\
  A_k &= \diag(I + D^0_k, I + D^1_k, \ldots, I + D^N_k),
\end{align}
$D^n_k$ is defined by~\eqref{eq:block-D} and $I$ is the $2\times 2$
identity matrix.  For the matrix $B$, we need to consider separately the
two cases corresponding to harmonic solutions where
\begin{gather}
  B_k = \left[ 
    \begin{array}{cc!{\colorvline}cc!{\colorvline}cc!{\colorvline}cc!{\colorvline}c}
	0 &  0 &  0 &  2 & &&&&  \\
	0 &  0 &  0 &  0 & &&&&  \\\colorhline
	0 & -1 &  0 &  0 &  0 & 1 &&&  \\
	1 &  0 &  0 &  0 & -1 & 0 &&&  \\\colorhline
	  &    &  0 & -1 &  0 & 0 &  0 & 1 & \\
	  &    &  1 &  0 &  0 & 0 & -1 & 0 & \\\colorhline
	  &    &    &    &  \ddots & & \ddots && \ddots
     \end{array}
  \right],
\end{gather}
and subharmonic solutions where
\begin{gather}
  B_k = \left[ 
    \begin{array}{cc!{\colorvline}cc!{\colorvline}cc!{\colorvline}cc!{\colorvline}c}
	0 &  1 &  0 &  1 & &&&& \\
	1 &  0 & -1 &  0 & &&&& \\\colorhline
	0 & -1 &  0 &  0 &  0 & 1 &&&  \\
	1 &  0 &  0 &  0 & -1 & 0 &&&  \\\colorhline
	  &    &  0 & -1 &  0 & 0 &  0 & 1 & \\
	  &    &  1 &  0 &  0 & 0 & -1 & 0 & \\\colorhline
	  &    &    &    &  \ddots  & & \ddots && \ddots
     \end{array}
  \right].
\end{gather}

\begin{figure}[bthp]
  \centering
  \includegraphics{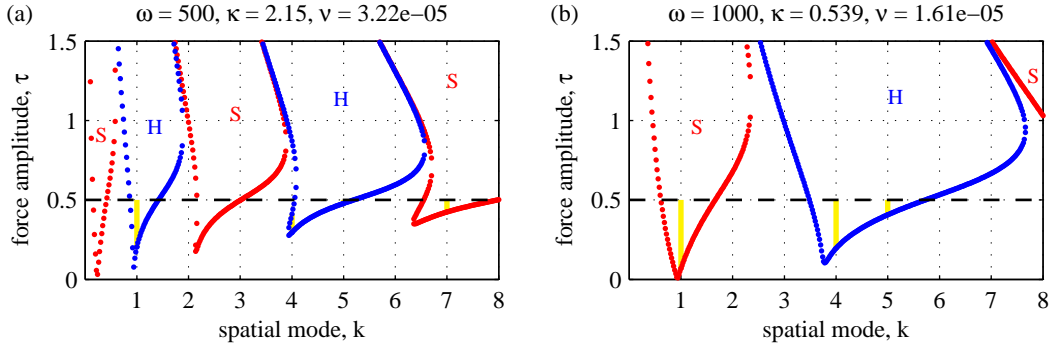}
  \caption{Ince-Strutt diagrams showing 
   \myrevisiontwo{the critical forcing amplitude $\tau$} for $\alpha=0$
    plotted against spatial wavenumber when $\omega =
    500\,\text{s}^{-1}$ (left) and $\omega = 1000\,\text{s}^{-1}$
    (right). The physically relevant parameters for instability are
    highlighted by the vertical solid lines.}
  \label{fig:eigs-lam0-omega}
\end{figure}
\begin{figure}[bthp]
  \centering
  \includegraphics{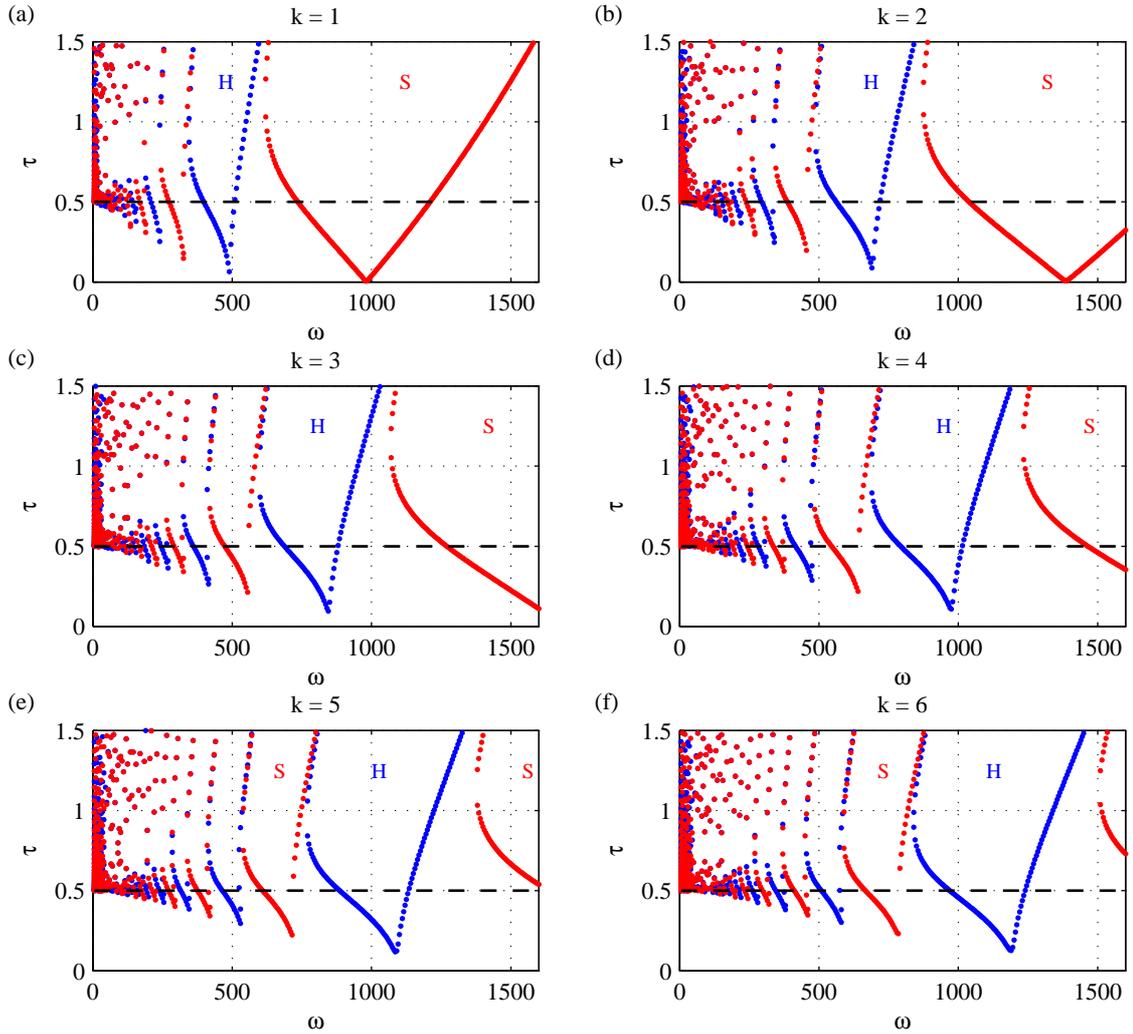}
  \caption{Ince-Strutt diagrams showing 
    \myrevisiontwo{the critical forcing amplitude $\tau$} for $\alpha=0$
    plotted against forcing frequency when $k$ varies from 1 to 6.}
  \label{fig:eigs-lam0-k}
\end{figure}

Figure~\ref{fig:eigs-lam0-omega} contains \emph{Ince-Strutt
  diagrams}~\cite{champneys-2009} 
  \myrevision{that depict $\tau$}
(corresponding to the critical forcing amplitude) plotted against the
spatial wavenumber $k$ for two values of the frequency
$\omega = 500$ and $1000\,\text{s}^{-1}$.
%
Each point on the diagram represents a periodic solution to the
linearized IB equations so that when taken together these points trace
out the marginal stability contours that divide parameter space into
regions corresponding to stable or unstable solutions.  The contours
have a characteristic ``tongue-'' or ``finger-like'' shape and alternate
between harmonic and subharmonic solutions, labelled ``H'' and ``S''
respectively.  Unstable solutions correspond to parameter values that
lie above and inside each tongue, while parameter values below the
tongues yield stable solutions.  Although we can expect parametric
resonance to occur for any choice of parameters located inside one of
the unstable tongues, the parameters are further constrained by the
following two physical considerations: first, that only modes
corresponding to integer values of $k$ are physically realizable; and
second, that the forcing amplitude must satisfy $\tau\leq \frac{1}{2}$
so that the BM stiffness $K(s,t)$ remains non-negative.  To get a
clearer idea of the unstable modes that correspond to physical BM
oscillations, we fix $k$ at integer values ranging from 1 through 6 and
display in Figure~\ref{fig:eigs-lam0-k} the stability plots as a
function of frequency $\omega$ on the horizontal axis.  In each case the
right-most tongue is labelled ``H'' or ``S'' and the subsequent tongues
moving to the left alternate between harmonic and subharmonic modes.  We
observe that as $k$ increases, the contours tend to widen and shift
downward and to the right; consequently, it is higher frequency
modes that are more susceptible to resonant instabilities.  Indeed, the
lowest wavenumber modes can be excited at very low values of the forcing
amplitude $\tau$ provided the forcing frequency is high enough.

To verify the existence of these resonant solutions from our linear
analysis in the case $\alpha=0$, we next perform numerical simulations
of the \myrevision{full governing IB equations
  \eqref{eq:dim-nse}--\eqref{eq:dXdt}, that now include nonlinearities
  from both the advection terms and the delta function integral terms}.
We use a standard approach similar to the algorithm described
in~\cite{peskin-2002}, in which the fluid variables are discretized on
an equally-spaced rectangular grid and the membrane on a moving set of
Lagrangian points.  A split-step projection method is used to solve the
incompressible Navier-Stokes equations and a regularized delta function
is used to approximate the integral terms that encapsulate the
fluid-structure interaction. We use the IB algorithm implemented in the
freely-available Matlab software package {\tt MatIB} whose
implementation details can be found in~\cite{froese-wiens-2013}.

\begin{figure}[bthp]
  \centering
  \includegraphics{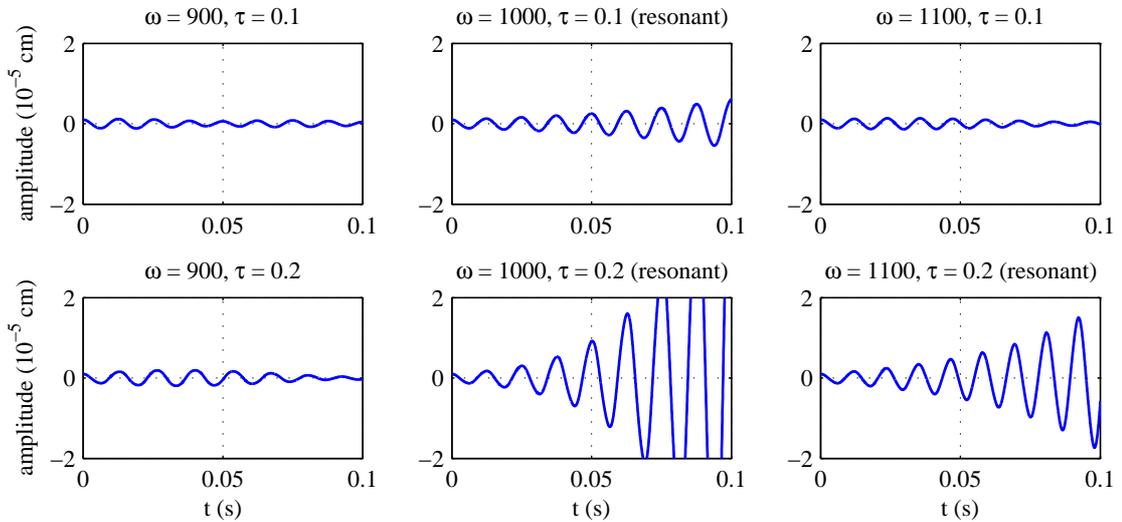}
  \caption{Time evolution of the IB peak amplitude when $\alpha=0$ given
    an initial $k=1$ mode cosine wave profile, for various values of
    parameters $\omega$ and $\tau$.}
  \label{fig:sims-lam0}
\end{figure}

Simulations are performed on a doubly-periodic fluid domain of size
$[-L,L]\times[-L,L]$ and we use the forcing parameters $\omega= 900,
1000, 1100\,\text{s}^{-1}$ and $\tau=0.1, 0.2$, with all other
parameters listed in Table~\ref{tab:params}.  Figure~\ref{fig:sims-lam0}
depicts the time evolution of the peak BM amplitude for an initial
membrane displacement corresponding to a $k=1$ cosine wave with
\myrevision{amplitude $10^{-6}$~cm}. For the parameter values chosen,
the results exhibit a range of behaviour including stable (non-resonant)
solutions in which the amplitude decays over time, as well as resonant
solutions that experience growth in amplitude by up to two orders of
magnitude for the largest values of forcing frequency $\omega$ and
amplitude $\tau$.
The Ince-Strutt diagram corresponding to $k=1$ from
Figure~\ref{fig:eigs-lam0-k} may be used to predict the solution
stability in these simulations, and the expected solution behaviour is
summarized in Table~\ref{tab:stability-lam0} for the parameter values
corresponding to the simulations.  Clearly, the linear analysis matches
the stability behaviour observed in simulations.  Most notably, for the
case of amplitude $\tau=0.1$ we capture for increasing $\omega$ how the
$k=1$ mode transitions from stable, through the marginal stability
boundary into an unstable tongue, and then returns again to the stable
region.

\begin{table}
  \caption{Analytical stability behaviour predicted for the parameters
    used in Figure~\ref{fig:sims-lam0} with $\alpha=0$.}
  \centering
  \footnotesize
  \renewcommand{\extrarowheight}{0.1cm}
  \begin{tabular}{c|ccc}\hline
    & $\omega = 900\,\text{s}^{-1}$ & $\omega = 1000\,\text{s}^{-1}$ &
    $\omega = 1100\,\text{s}^{-1}$\\ 
    \hline
    $\tau = 0.1$ & stable & unstable & stable  \\ 
    $\tau = 0.2$ & stable & unstable & unstable\\\hline
  \end{tabular}
  \label{tab:stability-lam0}
\end{table}

\subsection{Case $\mathbf{\alpha\ne 0}$}

\begin{figure}[bthp]
  \centering
  \includegraphics{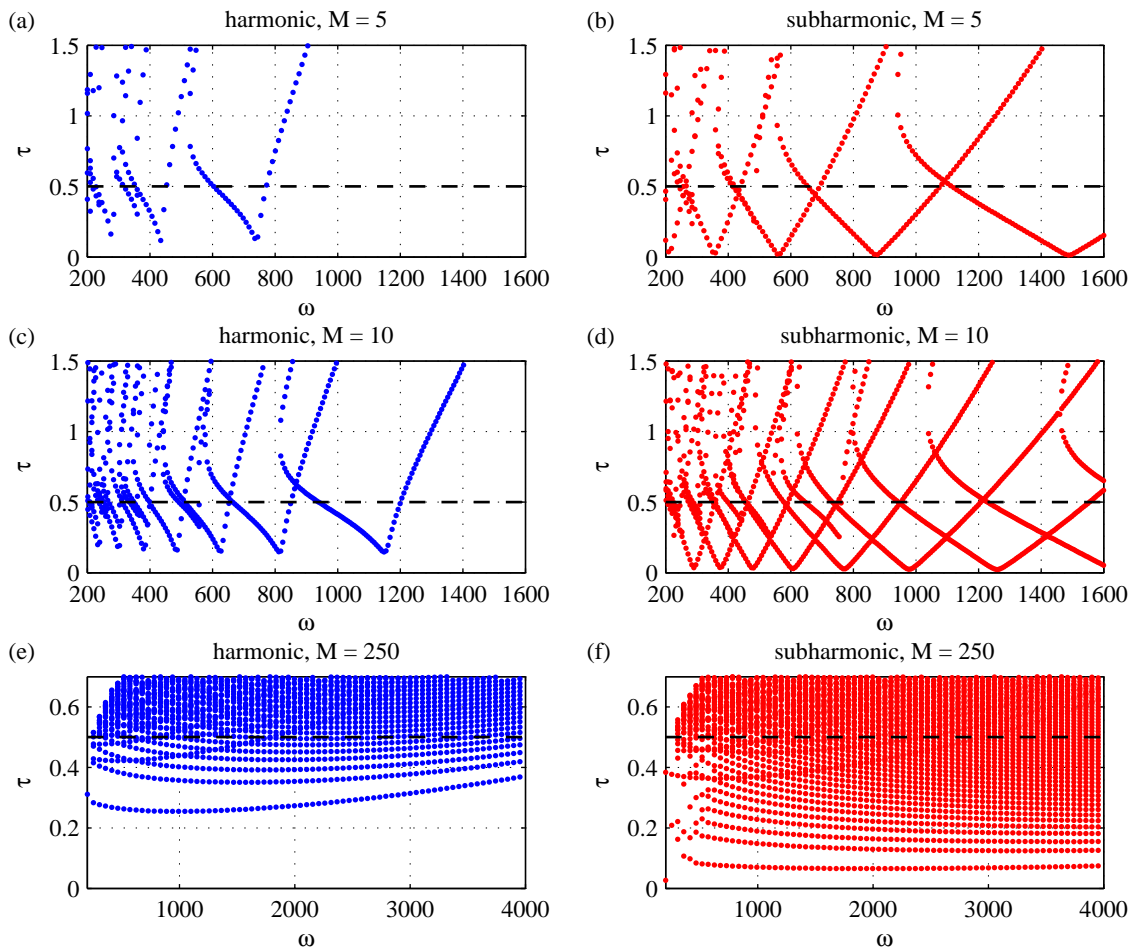}
  \caption{Ince-Strutt diagrams for the case $\alpha\ne 0$, depicting
    the convergence of 
    \myrevisiontwo{the critical forcing amplitude}
    as the number of spatial modes
    $M$ is increased from 5 (top) to 10 (middle) to 250 (bottom).  The
    eigenvalues are separated into harmonic (left) and subharmonic
    (right) modes. Note the different axes for the $M=250$ case.}
  \label{fig:eigs}
\end{figure}

We next investigate the stability of solutions in the spatially-coupled
case ($\alpha\ne 0$) where the BM stiffness varies exponentially along its length.
The stability contours are shown as plots of $\tau$ versus $\omega$ in
Figure~\ref{fig:eigs} using the same parameters listed in
Table~\ref{tab:params}.  Here we have displayed the harmonic and
subharmonic mode plots separately, and we also present two sets of
results that truncate the series solutions at different numbers of
spatial modes, $M=5$, 10 and 250 (in all cases using $N=20$ temporal
modes).  In contrast with the $\alpha=0$ results from the previous
section where stability contours are disjoint and the behaviour of a
given mode is easy to identify, we observe that contours overlap due to
the coupling between spatial modes.  A similar ``mode-mixing'' effect
has been observed in other physical systems such as the double
pendulum~\cite{jackel-mullin-1998}.  Furthermore, we find that the
number of stability contours depends strongly on the number of spatial
modes $M$ included in the truncated series expansion; in particular,
increasing $M$ gives rise to more stability contours that tend to pack more
closely together. As $M$ gets large, the contour ``tongue-tips'' sweep
out a smooth curve that divides parameter space into stable and
unstable solutions as seen in the bottom row of plots in
Figure~\ref{fig:eigs} for $M=250$. We note that convergence in the time
modes is much faster than in the spatial modes, thereby requiring that
$M$ be taken significantly larger than $N$ in order to achieve accurate
results.  \myrevision{We note in particular that when either $k > 250$
  or $n > 20$, all coefficients satisfy $|h_k^n| < 10^{-4}$ and so the
  neglected modes have negligible effect on our computed results.}


\myrevision{Although it is no longer possible to predict the growth or decay of a
single wavenumber mode as in the $\alpha = 0$ case, 
we can nevertheless still identify the region in parameter space 
corresponding to stable solutions.
Figures~\ref{fig:eigs}e and~\ref{fig:eigs}f show that for each forcing frequency
there is a value of $\tau_o$ in the interval
$[0,\frac{1}{2}]$ for which a forcing amplitude $0<\tau<\tau_o$ yields a
stable solution whereas amplitudes $\tau_o<\tau<\frac{1}{2}$ lead to
resonance.} Again, viscosity acts as a stabilizing mechanism in the
sense that increasing $\mu$ will increase $\tau_o$ and consequently
increase the size of the region in parameter space where solutions are
stable.  Varying $\sigma$ has the effect of changing the range of
resonant frequencies: increasing $\sigma$ causes the contours to spread
out, thereby increase the range of frequencies that result in unstable
solutions.


\begin{figure}[bthp]
  \centering
  \includegraphics{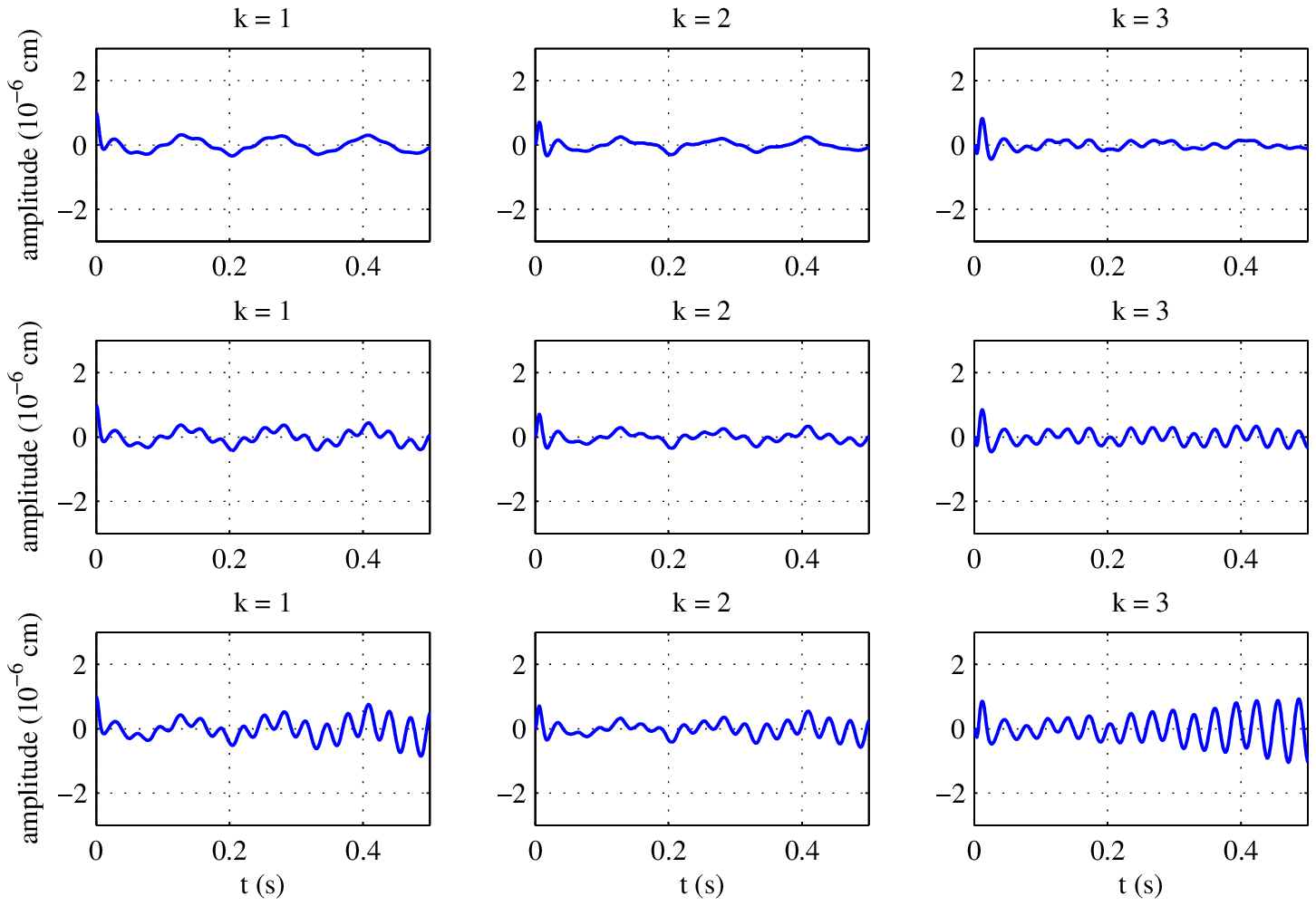}
  \caption{Time evolution of the amplitude of Fourier cosine
    coefficients from numerical simulations for internal forcing
    frequency $\omega = 400\,\text{s}^{-1}$ and stiffness forcing amplitude $\tau = 0.05, 
    0.08$ and 0.1 (top, middle, bottom).}
  \label{fig:sims-400}
\end{figure}

\begin{figure}[bthp]
  \centering
  \includegraphics{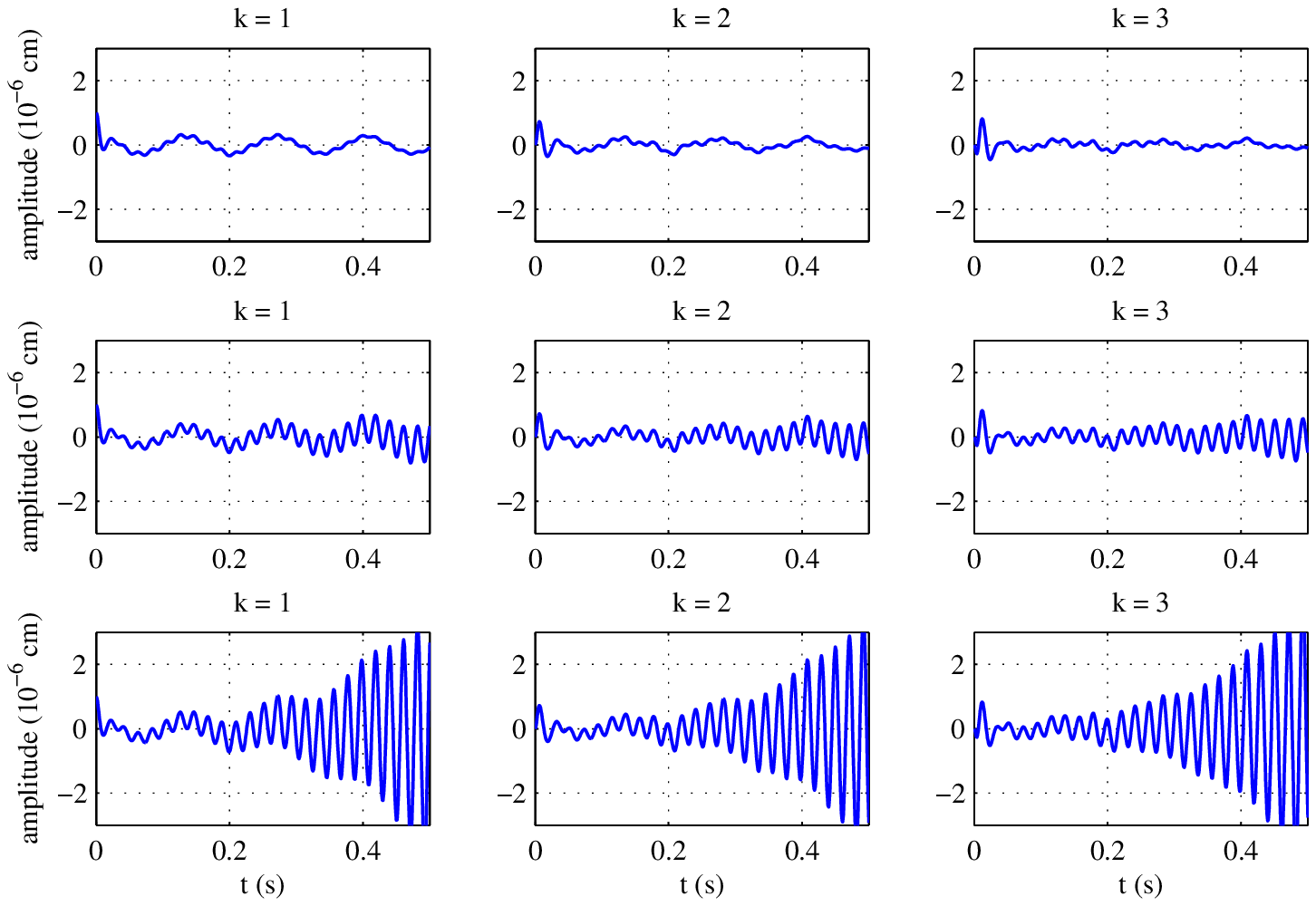}
  \caption{Time evolution of the amplitude of Fourier cosine
    coefficients from numerical simulations for internal forcing
    frequency $\omega = 600\,\text{s}^{-1}$ and stiffness forcing amplitude $\tau = 0.05,
    0.08$ and 0.1 (top, middle, bottom).}
  \label{fig:sims-600}
\end{figure}

\begin{figure}[bthp]
  \centering
  \includegraphics[width=\textwidth]{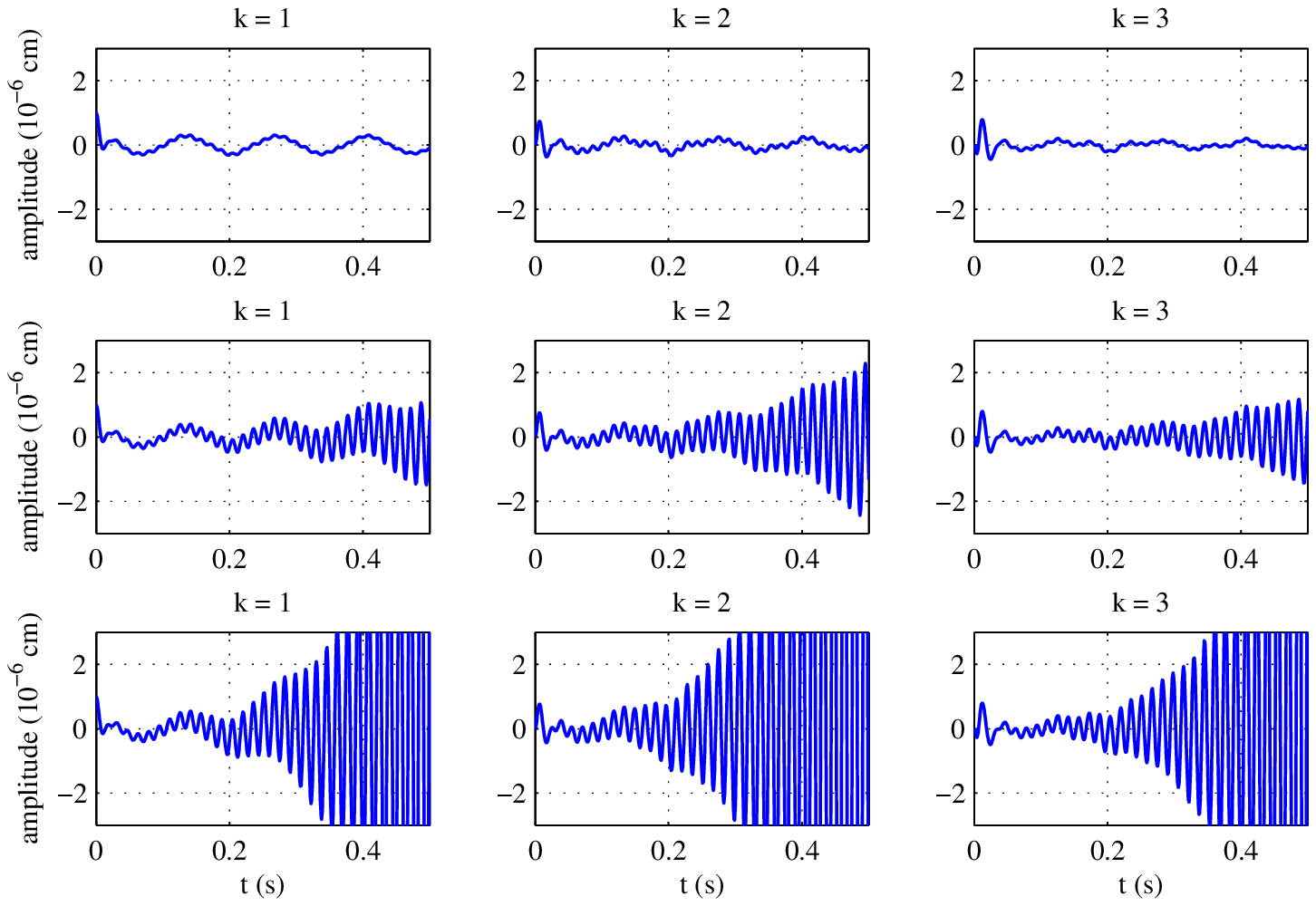}
  \caption{Time evolution of the amplitude of Fourier cosine
    coefficients from numerical simulations for internal forcing
    frequency $\omega = 800\,\text{s}^{-1}$ and stiffness forcing amplitude $\tau = 0.05,
    0.08$ and 0.1 (top, middle, bottom).}
  \label{fig:sims-800}
\end{figure}

Once again, we use numerical simulations of the full IB model equations
on the domain $\Omega = [-L,L]\times[-L,L]$ to validate the existence of
resonant modes found analytically.  For the initial membrane
configuration, we use the same cosine wave with 
\myrevision{amplitude $10^{-6}$~cm}
and wavenumber $k=1$.  Figures~\ref{fig:sims-400}, \ref{fig:sims-600}
and~\ref{fig:sims-800} display the time variation of the amplitude of
the first three Fourier cosine modes with forcing frequencies
$\omega=400, 600$ and $800\,\text{s}^{-1}$ respectively.  In each case
we also choose three different values of the forcing amplitude,
$\tau=0.05$, 0.08 and 0.1.  Even though the initial condition contains a
pure wavenumber $k=1$ mode, all $k$-modes are eventually excited because
of the mode-coupling that arises through the spatially-dependent
stiffness.  According to Figure~\ref{fig:eigs}, we expect the
$\tau=0.05$ cases to be stable (since the tips of all tongues lie above
this value of $\tau$) while taking $\tau$ any larger should destabilize
the solution.  Indeed, numerical simulations with $\tau = 0.05$ do show
that BM oscillations decay in time and that the parametric forcing is
insufficient to initiate an instability.  Furthermore, when $\tau$ is
increased to 0.08, the solutions become unstable, sustained oscillations
appear, and for the largest value of $\tau=0.1$ the peak amplitude grows
even larger.  It is important to note that in all of the resonant cases
simulated, the oscillation frequency is half of the internal forcing
frequency, which is a common signature of parametric resonance.

\myrevision{Slight differences arise from the fact that our numerical
  simulations are on a doubly-periodic domain of finite length, whereas
  the analysis assumes a fluid domain of infinite extent in $y$.
  Although we have chosen the domain size to be large enough that
  boundary effects are kept to a minimum, there are still interactions
  between periodic BM copies that cannot be completely eliminated in our
  simulations.}

\begin{figure}[bthp]
  \centering
  \includegraphics{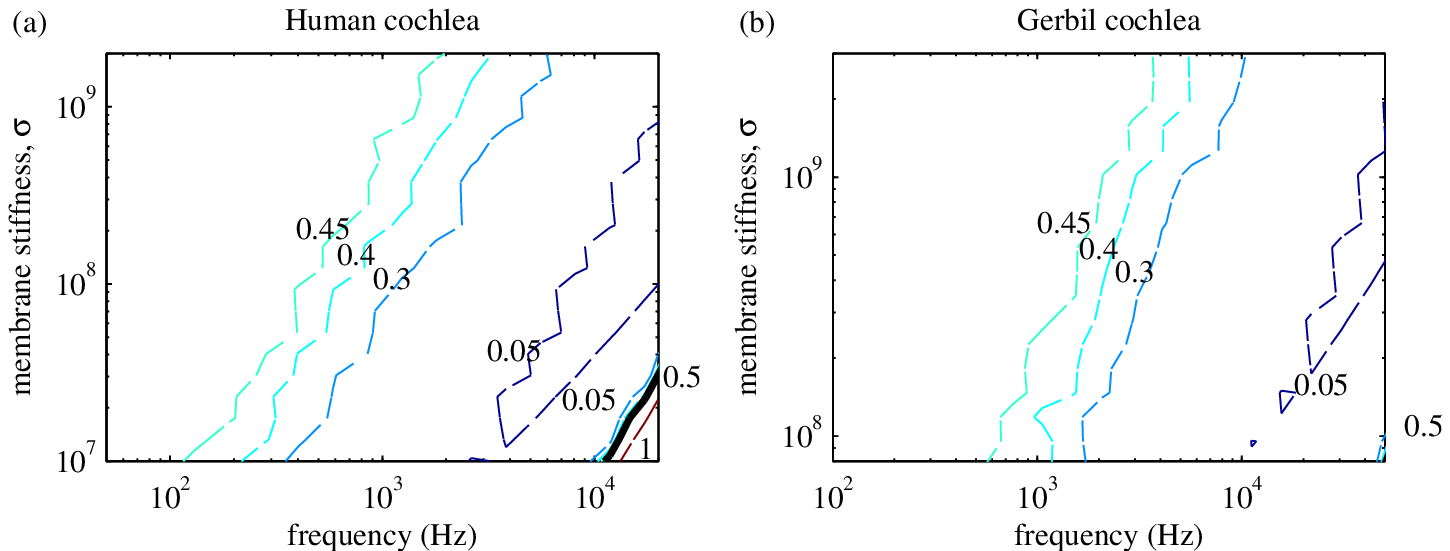}
  \caption{Contour plots of minimum $\tau$ required for parametric
    resonance with physically relevant parameters for the human cochlea (left)
    and the gerbil cochlea (right).
    The contour corresponding to $\tau_o=0.5$ is shown as a solid curve,
    and so only those parameters lying in the lower right corner of the
    diagram (small stiffness, large frequency) correspond to stable or
    non-resonant solutions.  
    }
  \label{fig:tau-contour}
\end{figure}

We conclude by discussing the existence of parametric resonances in our
cochlea model for the full range of physically-relevant BM parameters
(namely, $\sigma$).  Figure~\ref{fig:tau-contour} displays
contours of the \myrevision{smallest $\tau$ resulting in resonance}
predicted by our analysis using $N = 20$, $M = 250$ for a human cochlea
(left) and a gerbil cochlea (right).  The parameters for the human
cochlea are taken from Table~\ref{tab:params} except for the membrane
stiffness.  Experimental values for $\sigma$ reported in the literature
for human cochleas exhibit a large variation, ranging from $\sigma =
1\times 10^{7}\;\text{ g}\,\text{cm}^{-2}\,\text{s}^{-2}$ from
\cite{lesser-berkley-1972} to as high as $2\times
10^{9}\;\text{g}\,\text{cm}^{-2}\,\text{s}^{-2}$ in other
two-dimensional models (see~\cite[Table~1]{neely-1981}, for example).
We consider this entire range on the vertical axis of
Figure~\ref{fig:tau-contour}, while the horizontal axis extent covers to
the the entire range of audible frequencies $\omega\in[50, 20000]$~Hz
for humans.\ Stiffness values for the gerbil cochlea were extracted from
\cite[Fig.~8]{olson-etal-2012} where several experiments are summarized
and show $\sigma \in [1.5\times 10^8,2.0\times
10^9]$~$\text{g}\,\text{cm}^{-2}\,\text{s}^{-2}$ (a wider range is
plotted).  The stiffness decay parameter for the gerbil BM is estimated
to be $\lambda = 3.7 \text{cm}^{-1}$ and the length is $L=1.3$ cm
\cite{naidu-mountain-2007}.  The horizontal axis of the right plot
covers the audible range for a gerbil $\omega \in [100,50000]$~Hz.
From these plots, we observe that parametric resonance is possible ($\tau
< \frac{1}{2}$) for nearly all parameter values corresponding to the human
cochlea, except at the highest frequencies and the lowest values of
$\sigma$.  Therefore, we can conclude that parametric resonance arising
from fluid-structure interaction effects is possible in our cochlea
model for most sounds in the human audible range.
\myrevisiontwo{
A similar conclusion can be drawn for the gerbil cochlea.
A possible experiment to further explore these parameter ranges
is to study BM response to in vitro electrical stimulation of the 
cochlea~\cite{chan-hudspeth-2005,nakajima-etal-1994} wherein
electrically evoked otoacoustic emissions could be modeled
as a parametric forcing in the outer hair cells.
}

\section{Summary and Conclusions}

\leavethisout{
  \begin{myquestion}
    Is there any way to make the case that the resonant amplitude we
    observe is close to the actual value coming from the mechanical
    resonator models of the BM?  If not, it's not a problem, but perhaps
    we can point to future work in this area that will lead to the PNAS
    paper.
  \end{myquestion}
}

A immersed boundary model was developed for the basilar membrane in the
cochlea,
\myrevision{for the purpose of investigating the relevance of parametric
  resonance as a novel mechanism for amplification of BM oscillations.}
Our model captures the fluid-structure interaction that occurs between
the basilar membrane and the surrounding cochlear fluid.  Our work is
based upon a previous model from~\cite{leveque-peskin-lax-1988}, but
includes the additional effects of internal (parametric) forcing due to
variations in the elastic properties of the BM.
\myrevision{ The prime motivation for introducing such a parametric
  forcing derives from the work of Mammano and
  Ashmore~\cite{mammano-ashmore-1993} who have uncovered experimental
  evidence that oscillations of the outer hair cells embedded in the CP
  can lead to periodic modulation of the BM elastic stiffness.}

We demonstrated first that a parametrically-forced membrane can produce
travelling wave solutions that are similar to those observed in
\cite{leveque-peskin-lax-1988} for a passive BM.  A Floquet stability
analysis was then used to demonstrate the existence of resonant
solutions in the linearized IB equations.  The results were presented as
plots of the marginal stability contours in $\tau$-$\omega$ (forcing
amplitude-frequency) parameter space.  For a spatially homogeneous
membrane with a constant value of stiffness, the stability contours in
the corresponding Ince-Strutt diagram are disjoint for each Fourier
mode.  However, for the realistic case of a BM stiffness that varies
exponentially along its length, there is a mode-mixing effect in which
the stability contours overlap in parameter space.  We conclude that
internal forcing through via the BM stiffness at sufficiently large
amplitudes can induce parametric instability for any frequency in the
physiological range of human hearing.  These existence of these
resonances is verified using numerical simulations of a full
two-dimensional immersed boundary model of the cochlea.

Our main conclusion is that parametric resonances arising from
fluid-structure interactions in the cochlea are worthy of further study
as a possible contributing factor in the sound amplification ability of
human and other mammalian hearing systems.  One obvious focus for future
research is to develop a more complete cochlea model that couples the
fluid-structure interaction effects (giving rise to parametric
resonance) along with an existing model for BM mechanical
amplification~\cite{mammano-nobili-1993, markin-hudspeth-1995,
  neely-1981, ramamoorthy-deo-grosh-2007, reichenbach-hudspeth-2010},
which would thereby permit a comparison of the relative importance of
the combined effects.  \myrevision{This would also permit us to replace
  the BM stiffness parameter \eqref{eq:Kst} with a more physiologically
  relevant (non-separable) function in which the spatial dependence is
  determined by an existing mechanical model that has been validated
  against experiments.}  Our immersed boundary model is also an ideal
framework to investigate effects such as longitudinal coupling within
the BM~\cite{allen-sondhi-1979, kim-xin-2005, naidu-mountain-2001} or
bending-resistant stiffness that some studies claim are
important~\cite{allen-sondhi-1979, pozrikidis-2008}.  Finally,
\myrevision{the active process in the cochlea has been connected by some
  authors with \emph{spontaneous otoacoustic
    emissions}~\cite{markin-hudspeth-1995} for which a study of the
  potential impact of parametric resonance on amplifying such
  spontaneous oscillations would be a fascinating topic for future
  investigation.}
 
\leavethisout{
\begin{myquestion}
  \begin{itemize}
  \item Coupling along the BM in the longitudinal ($x$)
    direction~\cite{allen-sondhi-1979, kim-xin-2005,
      naidu-mountain-2001} but discounted by 
    \cite{jaffer-kunov-wong-2002}.   Bending-resistant beam model
    \cite{allen-sondhi-1979, pozrikidis-2008}.
  \item Oscillatory flow in tunnel of Corti can introduce a
    fluid-mediated longitudinal
    coupling~\cite{karavitaki-mountain-2007}.
  \item Do we say anything here about possible connections with
    otoacoustic emissions? 
  \end{itemize}
\end{myquestion}
}

\section*{Acknowledgments}

We would like to express our thanks to Brittany Froese and Jeffrey Wiens
for the use of their two-dimensional immersed boundary code
MatIB~\cite{froese-wiens-2013}.


\end{document}